\documentclass[superscriptaddress,aps,preprintnumbers,amsmath,amssymb,prd,nofootinbib,preprint]{revtex4-1}
\pdfoutput=1
\usepackage{graphicx}
\usepackage{epstopdf}
\usepackage{mathrsfs}
\usepackage{amssymb}
\usepackage{verbatim}
\usepackage{color}
\usepackage{multirow}
\usepackage{tcolorbox}
\usepackage{hyperref}

\numberwithin{equation}{section}

\makeatletter
\renewcommand{\p@subsection}{}
\renewcommand{\p@subsubsection}{}
\makeatother

\newcommand{\beq}{\begin{equation}}
\newcommand{\eeq}{\end{equation}}
\newcommand{\ga}{\lower.7ex\hbox{$\;\stackrel{\textstyle>}{\sim}\;$}}
\newcommand{\la}{\lower.7ex\hbox{$\;\stackrel{\textstyle<}{\sim}\;$}}

\usepackage{hyperref}
\usepackage{amsmath,bm}
\usepackage{physics}

\hypersetup{
    colorlinks = true,
    citecolor = {blue},
    linkcolor = {blue},
    urlcolor = {blue},
}

\begin{document}

\vspace{0.5cm}
\title{\makebox[\linewidth][c]{Axion Kinetic Misalignment and Parametric Resonance from Inflation}}

\author{Raymond T. Co}
\affiliation{Leinweber Center for Theoretical Physics, Department of Physics, University of Michigan, Ann Arbor, MI 48109, USA}
\author{Lawrence J. Hall}
\affiliation{Department of Physics, University of California, Berkeley, CA 94720, USA}
\author{Keisuke~Harigaya}
\affiliation{School of Natural Sciences, Institute for Advanced Study, Princeton, NJ 08540, USA}
\author{Keith~A.~Olive}
\affiliation{William I. Fine Theoretical Physics Institute, School of Physics and Astronomy, University of Minnesota, Minneapolis, MN 55455, USA}
\author{Sarunas~Verner}
\affiliation{William I. Fine Theoretical Physics Institute, School of Physics and Astronomy, University of Minnesota, Minneapolis, MN 55455, USA}

\begin{abstract}
Axion cold dark matter from standard misalignment typically requires a decay constant $f_a~\gtrsim~10^{11}$~GeV. Kinetic misalignment and parametric resonance easily allow lower values of $f_a$ when the radial Peccei-Quinn (PQ) symmetry breaking field takes large initial values. Here, we consider the effects of inflation on kinetic misalignment and parametric resonance. We assume that the initial PQ field value is determined by quantum fluctuations, and is set by the Hubble parameter during inflation, $H_I$, and the PQ field mass. PQ field oscillations begin before or after the completion of reheating after inflation at a temperature $T_R$. We determine the range of $f_a$ and the inflationary parameters $(H_I, T_R)$ consistent with axion dark matter for a quartic potential for the PQ field. We find that $4\times 10^8$ GeV $< f_a < 10^{11}$ GeV can consistently produce axion dark matter. A significant portion of the allowed parameter space predicts rare kaon decays, $K_L \rightarrow (\pi^0 + \rm{missing \; energy})$, and/or suppression of structure formation on small scales.
\begin{center}
{\tt LCTP-20-06, UMN-TH-3912/20, FTPI-MINN-20/02} 
\end{center}

\end{abstract}

\maketitle

\begingroup
\hypersetup{linkcolor=black}
\renewcommand{\baselinestretch}{1.2}\normalsize
\tableofcontents
\renewcommand{\baselinestretch}{2}\normalsize
\endgroup

\newpage

\section{Introduction}
Why is the CP symmetry badly broken by the weak interaction, but preserved by the strong interaction~\cite{thooft1, thooft2, neut1, NEDM_EX}? The Peccei-Quinn~(PQ) mechanism~\cite{pq1} remains a highly plausible resolution to this problem, replacing the static CP-violating vacuum angle $\theta$ by a dynamical field $\theta(x) = a(x)/f_{a} $, where $a(x)$ is the pseudo Nambu-Goldstone boson known as the axion~\cite{weinberg, wilczek}, and $f_a$ is the PQ symmetry breaking scale. We consider a scenario in which the PQ symmetry is broken during inflation. When a global~$U(1)_{\rm PQ}$ symmetry is spontaneously broken, the axion arises as a Nambu-Goldstone boson. As the temperature approaches the QCD confinement scale $\Lambda_{\rm QCD}$, the explicit PQ symmetry breaking by the QCD anomaly becomes effective, causing the axion field to obtain a non-negligible mass and coherently oscillate around the minimum. The resulting oscillation energy accounts for the cold dark matter abundance~\cite{cosm1} and this axion production is known as the misalignment mechanism.

The key parameter determining the axion relic density is the axion decay constant, $f_a$, which is related to the axion mass~\cite{weinberg}
\begin{equation}
m_a \simeq 6 \, \text{eV} \left(\frac{10^6 \, \text{GeV}}{f_a} \right).
\label{mafa}
\end{equation}
Astrophysical constraints provide a lower bound on the decay constant,  $f_a \gtrsim \mathcal{O}(10^{7} - 10^{8}) \, \text{GeV}$~\cite{astr1, astr2}. 
For $f_a \simeq 10^{12}$ GeV, the correct relic density $\Omega_a h^2 \simeq 0.12$~\cite{planck18} is obtained for initial field values $\theta_i$ of order unity. For $f_a \gg 10^{12}$ GeV, the initial misalignment angle must be tuned so that $f_a \theta_i^2$ stays approximately constant. This fine-tuning can be relaxed if the Universe experiences
some late-time matter domination \cite{kmy} or if inflationary dynamics relaxes the initial misalignment~\cite{Dimopoulos:1988pw,Dvali:1995ce}. For $f_a \ll 10^{12}$ GeV, the axion tends to be underproduced unless $\theta_i$ is very close to the hilltop of the potential, where $\theta_i = \pi$, or a different production mechanism is in effect.

This standard understanding, however, may be incorrect. The axion field is the angular component of a complex scalar field, $P= S e^{i \theta}/\sqrt{2}$, and the cosmological history of the axion may in fact be affected if there is a large initial value for the radial component, $S$, which we call the saxion.
As the saxion starts to oscillate around the origin, axions are produced via parametric resonance (PR)~\cite{parres1,parres2,parres3}, which is very efficient and can yield axion dark matter for values of $f_a$ much lower than in conventional misalignment~\cite{parres2,parres3}. PR produced axions may have non-negligible velocity and affect structure formation in the Universe.

A large initial saxion field may probe higher dimension operators in the potential that explicitly break the PQ symmetry, and therefore depend on the angular component $\theta$. In this case, as the field starts to oscillate, these operators impart a kick generating a large angular speed $\dot{\theta} \equiv d\theta/dt$. Much later in its evolution, the resulting kinetic energy of the axion field may dominate over the QCD-generated axion potential, delaying the usual axion field oscillations. This kinetic misalignment mechanism (KMM)~\cite{chh} can explain axion dark matter for small values of $f_a$\footnote{A non-zero kinetic energy of the axion field can also explain the baryon asymmetry of the Universe through a mechanism called axiogenesis~\cite{axiogen}.}.
It is possible to suppress the axion abundance at large $f_a$ with a fine-tuned initial condition for $(\theta, \dot\theta)$~\cite{Chang:2019tvx}.

In both PR and KMM mechanisms, in addition to $f_a$, the final axion abundance depends on the potential (including the PQ-breaking term that induces the angular kick for KMM) and the initial position of the radial component, $S_i$, but does not depend sensitively on the initial condition for the angular component.  The PQ symmetry is broken before inflation, and in \cite{parres2,chh} it is assumed that the large $S_i$ is set
independent of the dynamics of inflation.

In this paper we study PR and KMM in detail. We find that KMM is not always separable from PR.
In KMM, unless the rotation of $P$ is sufficiently circular, PR also occurs. The axion abundance is generically given by the sum of both contributions. The PR contribution is removed if the produced axion fluctuations are thermalized. The KMM contribution is protected by the conservation of the PQ charge and is not removed by thermalization~\cite{axiogen}.

We investigate PR and KMM when it is intimately linked to the inflationary era.  First, we assume that the initial saxion field value, $S_i$, is set by quantum fluctuations during inflation, replacing an unknown initial condition with dynamics involving the Hubble parameter during inflation, $H_I$.
We assume no specific model of inflation, but take Starobinsky-like inflation \cite{Staro} as a motivated and predictive example that relates the inflaton mass and the Hubble parameter, with the predicted values for these from the normalization of CMB anisotropies.
Second, we explore the possibility that the saxion field oscillations begin before or after the completion of reheating. For the former, the axion abundance depends on the reheat temperature, $T_R$. With these connections between PR, KMM, and inflation, we determine the range of the inflationary parameters $(H_I, T_R)$ consistent with axion dark matter using a quartic potential for the PQ field.  Our results apply quite generally for any model of inflation, though numerical results can be sharpened in simple specific models.  

The conventional misalignment mechanism predicts the axion mass $m_a \lesssim \mathcal{O}\, (0.01) \, \rm{meV}$, while PR and KMM allow the axion mass to be higher, with $m_a \simeq \mathcal{O}(0.01 - 100) \, \rm{meV}$, a mass range that is being probed by various experimental investigations~\cite{expaxion}. Other possible mechanisms that predict heavier axions include:
1) unstable cosmic domain wall decays~\cite{domwall}, 2) anharmonicity effects when the axion misalignment angle $\theta_i$ approaches $\pi$ due to inflationary dynamics~\cite{anh}, and 3) axion production in non-standard cosmologies~\cite{kination}.

In what follows, we first consider the consequences and conditions for KMM and PR in Secs.~\ref{sec:kmm} and \ref{sec:PR}, respectively. We use a simple quartic potential for $|P|$.
We discuss the inflationary dynamics of $P$ in Sec.~\ref{sec:before}. During inflation, quantum fluctuations are used to derive the initial conditions for the saxion. 
The axion abundance in each case is
calculated in Sec.~\ref{sec:abund},
where we derive bounds on the axion decay constant. In KMM and PR, a large saxion abundance is also produced and may cause cosmological problems. Thermalization of the saxion condensate and the subsequent decay of massive saxions are discussed in Secs.~\ref{sec:therm} and \ref{sec:decay}.
Finally, a discussion and our conclusions are given in Sec.~\ref{sec:disc}.

\section{Kinetic Misalignment Mechanism with Quartic Potential}
\label{sec:kmm}

Once the temperature of the Universe drops down to the GeV scale, non-perturbative QCD effects generate a potential for the axion
\begin{equation}
\label{axpot}
    V = m_{a}(T)^2 f_a^{2} \left(1 - \cos{\frac{a}{f_a}} \right)\, .
\end{equation}
Above the QCD scale the axion mass $m_a(T)$ is a function of temperature. We adopt the dilute instanton gas approximation~\cite{Gross:1980br},
\begin{equation}
\label{eq:maT}
m_a(T) \simeq m_a \left(\frac{\Lambda_{\rm QCD}}{T} \right)^4,
\end{equation}
where $m_a$ is given in Eq.~(\ref{mafa}) and we assume that the QCD phase transition occurs at~$\Lambda_{\rm QCD} \simeq 150 \, \rm{MeV}$.

If $\dot{\theta}$ is sufficiently small, the axion field starts to oscillate in this potential at a temperature $T_*$, given by $3H(T_*) = m_a(T_*)$, leading to the conventional misalignment mechanism~\cite{cosm1}.  If, on the other hand, at $T_*$ the kinetic energy density is larger than the potential barrier, or equivalently $\dot{\theta}(T_*) > 2 m_a(T_*)$, then kinetic misalignment occurs and the axion oscillations are delayed until the kinetic energy is below that of the potential~\cite{chh}. This condition corresponds to a yield of the PQ charge greater than a critical value
\begin{equation}
\label{ytheta}
Y_{\theta} \equiv \frac{n_{\theta}}{s} > Y_{\rm crit} \simeq 0.11 \left(\frac{f_a}{10^9 \, \text{GeV}} \right)^{13/6},
\end{equation}
where $s = \frac{2 \pi^2}{45} g_* T_*^3$ and $Y_{\rm crit}$ is the minimal yield for kinetic misalignment, and is evaluated  using $g_*(T_*) \simeq 26$, see, for example,~\cite{bors}. The PQ charge density $n_{\theta}$ will be defined below. A detailed calculation of $Y_{\rm crit}$ is given in Appendix~\ref{app:crit}.

Taking $Y_{\theta} \gg Y_{\rm crit}$, the energy density of oscillating axion is given by~\cite{chh}
\begin{equation}
\label{rho_KMM}
\frac{\rho_{a}}{s} \simeq 2 m_a Y_\theta \, ,
\end{equation}
leading to
\begin{equation}
\label{ab1}
\Omega_a h^2 \simeq 3 \times 10^{-3} \, Y_{\theta} \left(\frac{10^{9} \, \text{GeV}}{f_a} \right).
\end{equation}
Requiring the observed amount of dark matter $\Omega_a h^2 \simeq 0.12$~\cite{planck18}, we can use Eq.~(\ref{ytheta}) to obtain a critical value $f_a^{\rm crit} \simeq 1.5 \times 10^{11} \, \text{GeV}$, above which the conventional misalignment mechanism occurs. Therefore, in this work we restrict our attention to
\begin{equation}
\label{eq:fa_KMM}
f_a \lesssim 1.5 \times 10^{11} \, \text{GeV} \, ,
\end{equation}
where the kinetic misalignment mechanism accounts for the correct dark matter abundance.

To generate a large $Y_\theta$ at a much earlier epoch, we consider a complex scalar field, $P$, that transforms under the PQ symmetry, $P \rightarrow e^{i \alpha} P$, and has a potential
\beq
V = V_0 + V_{A} \, ,
\eeq
where the PQ-conserving term contains a quartic coupling
\begin{equation}
\label{pot}
V_0 = \lambda^2 \left(|P|^2 - \frac{f_{a}^2}{2} \right)^2.
\end{equation}
We also introduce a PQ-violating, higher-dimensional potential term, 
\beq
\label{potA}
V_{A} = \frac{A P^n}{n M_{\rm Pl}^{n-3}} + \text{h.c.}\, ,
\eeq
where $A$ is a dimensionful coupling and $M_{\rm Pl}\simeq 2.4 \times 10^{18} \, \rm{GeV}$ is the reduced Planck mass. Such terms are plausible since the PQ symmetry is explicitly broken by the QCD anomaly; if the PQ symmetry arises as an accidental symmetry from other exact symmetries~\cite{accidentalPQ}, explicit PQ symmetry breaking by higher dimensional operators is expected. Violation of global symmetries by quantum gravity effects~\cite{quantumgravity} may also generate such terms.

We parameterize the complex field $P$ by
\begin{equation}
\label{pfield1}
P \equiv \frac{S}{\sqrt{2}}e^{i \theta}\, ,
\end{equation}
where $S$ is the saxion (radial) component and $\theta = a(x)/f_a$ is the axion (angular) component. We assume that the saxion takes a large field value during inflation, e.g., due to quantum fluctuations as we discuss in more detail in Sec.~\ref{subsec:saxion}. After inflation, as the Hubble parameter drops to the saxion mass $m_{S_i}$ around the field value $S_i$, the field $P$ starts to oscillate at time $t_i$, where $3 H(t_i) \simeq m_{S_i}$.   Any motion in the angular direction leads to a PQ charge density
\begin{equation}
\label{numden1}
n_{\theta} \; = \; i \dot{P}^* P - i \dot{P} P^* \;=\; S^2 \dot{\theta}\, ,
\end{equation}
which is the Noether charge density associated with the axion shift symmetry.
The rotation is initiated in the same manner as Affleck-Dine baryogenesis~\cite{Affleck:1984fy}. The PQ-violating potential (\ref{potA}) imparts a kick in the $\theta$ (angular) direction, resulting in an initial field rotation, $\dot{\theta_i} \neq 0$, giving (see Appendix~\ref{app:asym})
\begin{equation}
\label{numden2}
n_{\theta_i} \; = \; \frac{1}{m_{S_i}} \left(\frac{i A}{M_{\rm Pl}^{n-3}}\right) (P^{n*} -  P^{n}) \;=\; \epsilon \frac{V_0(P_i)}{m_{S_i}}\, ,
\end{equation}
where the last equality defines $\epsilon$.
The parameter $\epsilon$ characterizes the shape of the motion and is determined by the angular potential gradient relative to the radial potential gradient. 
As we discuss in more detail in Sec.~\ref{sec:PR}, 
for values of $\epsilon \lesssim 0.8$, 
axion production through parametric resonance must be considered.

\begin{figure}[!ht]
\centering
\includegraphics[width=0.75\linewidth]{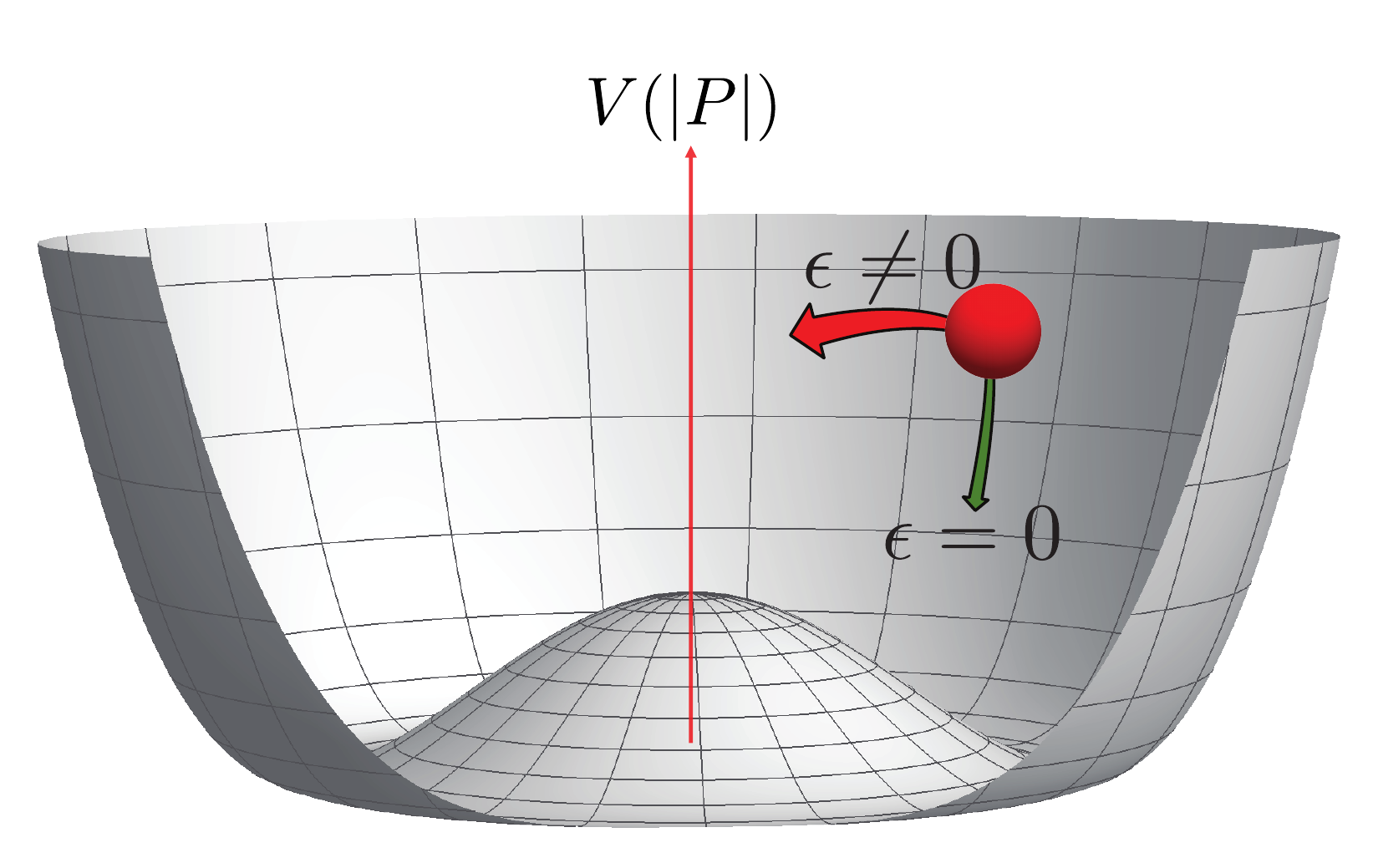}
\caption{The motion of a complex scalar field $P$. It remains frozen (overdamped) until the
Hubble scale becomes comparable to the initial saxion mass, i.e., $m_{S_i} t_i \sim 1$. 
The parameter $\epsilon$ characterizes the shape of the motion at the onset of radial oscillation/angular rotation; $\epsilon = 0$ and $1$ correspond to purely radial and angular motion, respectively.}
\label{mexhat1}
\end{figure}

Fig.~\ref{mexhat1} illustrates the motion of a complex scalar field $P$ along the quartic potential~(\ref{pot}).
Due to the friction from cosmic expansion, the radius of the rotation of $P$ decreases until $P$ reaches the bottom of the potential, on which $P$ continues to rotate. If the rotation is fast enough, i.e.~the PQ charge is large enough, kinetic misalignment is at work.

\section{Parametric Resonance with Quartic Potential}
\label{sec:PR}
If the motion of the PQ symmetry breaking field is not sufficiently circular, the value of $|P|$ changes rapidly during the cycle. This results in parametric resonance~\cite{parres1}, where the energy density of the coherent condensate is transferred into that of excited modes. In this section, we review the consequences of parametric resonance (PR) and derive the condition on $\epsilon$ for PR to occur.

The field $P$ oscillates at the frequency $\lambda S_i$ and, due to the quartic coupling, contributes an oscillatory effective mass to $\chi$, defined as the
excitation modes around the coherent motion of $P$.
An exponential enhancement of $\chi$ occurs in a momentum mode $k_\chi$ when the two frequencies, $\lambda S_i$ and $k_\chi$, satisfy a resonance condition. The initial energy density, stored in the coherent condensate of $S$, is transferred into the resonant modes of $\chi$. Such an efficient transfer begins shortly after the onset of $S$ oscillations and ceases when back-reactions become sufficiently strong. The back-reactions of the produced $\chi$ include scattering with the $S$ coherent condensate and changes in the effective mass of $\chi$ through self-scattering~\cite{parres1}. 
At this stage, PR has produced comparable number densities of axions and saxions
\begin{equation}
\label{axion_PR}
n_{a_i} \sim n_{S_i} \sim \frac{V_0(P_i)}{m_{S_i}}\, ,
\end{equation}
and the energy of $P$ is now predominantly in the excited modes and in the rotating mode\footnote{Due to conservation of PQ charge in the form of rotation, the energy density stored in rotation will remain even though the back-reaction results in transferring energy from the coherent condensate to fluctuations.}. The Peccei-Quinn symmetry is restored non-thermally due to the large fluctuations of $P$. The axions produced by PR can also contribute to the dark matter abundance~\cite{parres2} and are constrained from the warmness of dark matter, as discussed in Sec.~\ref{sec:PR_DM}.

PR is effective only when the adiabatic condition $\dot{m}_S/m_S^2 < 1$ is violated. When $P$ starts to oscillate, $S_i \gg f_a$ and $m_S(S) = \sqrt{3} \, \lambda S$. This implies that $\dot{m}_S/m_S^2 = \dot{S} / \sqrt{3} \, \lambda S^2$, and is maximized when $S$ is at its minimum value during a cycle where $\dot{S}$ also takes the maximum value $\dot{S}_{\rm max}$. The maximum kinetic energy $\dot{S}_{\rm max}^2/2$ during a cycle originates from the potential energy associated with the radial oscillation $(1 - \epsilon) \lambda^2 S_i^4 / 4$, which gives $\dot{S}_{\rm max} \simeq \sqrt{1 - \epsilon} \lambda S_i^2 / \sqrt{2}$. The minimum value of $S$ during a cycle is obtained from conservation of both energy and PQ charge, and is given by $S_{\rm min} = \epsilon S_i$. As a result, the adiabatic factor takes the following maximum value during a cycle
\begin{equation}
\left. \frac{\dot{m}_S}{m_S^2} \right|_{\max} \simeq \frac{\sqrt{1 - \epsilon}}{\sqrt{6} \, \epsilon^2} \, ,
\end{equation}
independent of $S_i$.
This rough estimate therefore shows that the adiabatic condition is violated when $\epsilon \lesssim 0.5$. We performed a numerical calculation for the two-field dynamics and found a critical value
\begin{equation}
\label{ep_crit}
\epsilon_{\rm crit} \simeq 0.8 ,
\end{equation}
below which PR occurs.

\section{Inflation and the Onset of Radial Oscillations}
\label{sec:before}

In this section, we study the saxion dynamics in the context of inflation and derive constraints relevant to axion dark matter from kinetic misalignment or parametric resonance. 

\subsection{Inflaton Dynamics}
\label{subsec:inflaton}

After the de Sitter phase of inflation with the Hubble parameter $H_I$,
the Universe becomes dominated by oscillations of the inflaton field $\phi$, and the expansion rate is that of a matter-dominated Universe, i.e., $\rho_{\phi} \sim R^{-3}$, with $R$ the scale factor. The energy of the inflaton oscillations is transferred into that of radiation,  in the process called reheating. After reheating completes at a temperature $T_R$, a radiation-dominated Universe begins.

While our general results do not depend on the details
of reheating, when we relate the reheating 
temperature $T_R$ to the inflaton decay rate, we assume that reheating after inflation occurs by the perturbative decay of the inflaton with a decay rate $\Gamma_\phi$. The transition from the matter-dominated era to the radiation-dominated era occurs when 
$\Gamma_{\phi} \, t \simeq 1$ or, equivalently, when $\Gamma_{\phi} \, \simeq 3 H /2$.
The reheating temperature can be expressed as
\begin{equation}
\label{reh}
T_R \simeq \left( \frac{40}{\pi^2 g_*} \right)^{1/4} \sqrt{M_{\rm Pl} \Gamma_{\phi}}\, ,
\end{equation}
where the effective number of degrees of freedom for the Standard Model is given by $g_* = 106.75$ (or higher in any extension of the Standard Model). We assume the full Standard Model value throughout this paper unless explicitely noted.

The decay rate of the inflaton field can be parameterized as
\begin{equation}
\label{decay2}
\Gamma_{\phi} = \frac{y^2}{8 \pi} m_{\phi}\, ,
\end{equation}
where $m_{\phi}$ is the inflaton mass.
The coupling $y$ is highly dependent on how the inflaton couples the Standard Model sector. Combining
Eqs.~(\ref{reh}) and (\ref{decay2}), we can write
\begin{equation}
\label{ycoup}
y = \left( \frac{\pi^2 g_*}{90} \right)^{1/4} \frac{2 \sqrt{3\pi} \, T_R}{\sqrt{m_\phi M_{\rm Pl}}} \simeq 11 \frac{T_R}{\sqrt{m_\phi M_{\rm Pl}}}\, .
\end{equation}

Our results
will be derived without the need to specify a particular model of inflation. We need only to specify the Hubble scale during inflation, $H_I$, and the reheat temperature, $T_R$. However, in places where we 
extract some numerical results, we will often assume the Starobinsky form for the inflationary potential
\begin{equation}
\label{staro}
V(\phi) =\frac{3}{4} m_{\phi}^2 M_{\rm Pl}^2 \left(1 - e^{-\sqrt{2/3} \phi/M_{\rm Pl}} \right)^2,
\end{equation}
where $\phi$ is the inflaton field and $m_{\phi}$ is the mass of the inflaton. The Hubble parameter during inflation is
\begin{equation}
H_I \simeq \sqrt{\frac{V(\phi)}{3M_{\rm Pl}^2}} \simeq \frac{m_{\phi}}{2}.
\end{equation}
The scale of inflation is
fixed from the normalization of CMB anisotropies. Using, $A_s = 2.1 \times 10^{-9}$ \cite{planck18}, and assuming 55 e-foldings of inflation, one finds
$m_{\phi} \simeq 3 \times 10^{13} \, \text{GeV}$ and $H_I \simeq 1.5 \times 10^{13} \, \text{GeV}$.

\subsection{Saxion Dynamics}
\label{subsec:saxion}
We assume that the initial value of the saxion field is set during inflation, when quantum fluctuations drive scalar expectation values of $\langle S^2 \rangle$ \cite{h4m2}. We take the saxion mass to be much less than the Hubble scale during inflation, $H_I$, so that quantum fluctuations grow as $H_I^3 t$ up to a fixed value given by
\begin{equation}
\label{sfield2}
\langle S^2 \rangle = \frac{3}{8 \pi^2} \frac{H_I^4}{m_S^2} \, \, .
\end{equation}
Long-wavelength modes of these fluctuations obey the same classical equations of motions
as $\langle S \rangle$, and, therefore, we may take (\ref{sfield2}) as an initial condition, $S_i$, for the saxion field and radial oscillations. 

Assuming $S_i \gg f_a$, at such large field values the saxion mass resulting from~(\ref{pot}) is
\begin{equation}
\label{radialmass}
m_S \simeq \sqrt{3} \lambda S \, . 
\end{equation}
If we use this mass with (\ref{sfield2}), we find a simple form for the initial saxion field
\begin{equation}
S_i \simeq \frac{H_I}{2^{3/4} \sqrt{\pi \lambda} } \, ,
\label{si}
\end{equation}
and combined with~(\ref{radialmass}), the initial saxion mass is
\begin{equation}
m_{S_i} \simeq  \sqrt{\frac{3 \lambda}{\pi}} \frac{H_I}{2^{3/4}}\, .
\end{equation}

Saxion oscillations begin at time $t_i$, where $m_{S_i} \sim 3H(t_i)$. We consider first the case where this occurs before reheating, namely where
\begin{equation}
\label{radialdec}
m_{S_i} \gtrsim 2 \, \Gamma_{\phi}\, .
\end{equation}
This requires
\beq
\lambda \gtrsim \frac{\sqrt{2} \pi^3 g_* }{15}  \frac{T_R^4}{H_I^2 M_{\rm Pl}^2} \simeq 5 \times 10^{-11}  \left(\frac{T_R}{10^{12} \, \text{GeV}}  \right)^4 \left(\frac{10^{12} \, \text{GeV}}{H_I}  \right)^2\, .
\label{boundary}
\eeq
We stress that Eq.~(\ref{boundary}) is (inflationary) model-independent.
If we use Eq.~(\ref{ycoup}), we obtain
\begin{equation}
\label{lambda2}
\lambda \gtrsim 8 \times 10^{-2} \, y^4 \, \, .
\end{equation}

When the constraint~(\ref{boundary}) is violated, saxion oscillations begin after reheating.
Using the standard expression $H^2 = \rho_R/3 M_{\rm Pl}^2$ in a radiation-dominated Universe, we obtain the following oscillation temperature
\begin{equation}
T_{i} \simeq  \left(\frac{15 \lambda}{\sqrt{2} g_* \pi^3} \right)^{1/4} \sqrt{H_I M_{\rm Pl}}\, ,
\label{ti}
\end{equation}
which does not depend on the reheating temperature, $T_R$.

Later, we fix $\lambda$ so that the axion population from PR or KMM explains the observed dark matter abundance.
If the required initial saxion field value is too large, a second period of inflation by the saxion potential energy occurs~\cite{go}. This means that, after fixing $\lambda$ to reproduce the observed dark matter abundance with the estimation assuming no second inflation, in the parameter region where the second inflation actually occurs, axions are always underproduced. 
Thus, we require $V_0(S_i)$ to be smaller than the total energy density of the universe when oscillation begins, or
\beq
\label{potcond}
\frac{\lambda^2S_i^4}{4} < 3 M_{\rm Pl}^2 H(t_i)^2 \, .
\eeq
Using $m_{S_i} \sim 3H({t_i})$, this is equivalent to the condition
\beq
S_i < 2 M_{\rm Pl} \, , 
\eeq
which leads to
\beq
\lambda > \frac{1}{8 \sqrt{2} \pi} \left(\frac{H_I}{M_{\rm Pl}} \right)^2 \simeq 5 \times 10^{-15} \left(\frac{H_I}{10^{12} \, \text{GeV}} \right)^2 \, .
\label{noinfl}
\eeq

For oscillations before reheating, we derive relations between the scale factors at various points of the cosmological history of the saxion, which are used later. Prior to reheating, the Hubble parameter is given by
\beq
H = H_I \left(\frac{R_\phi}{R}\right)^{3/2},
\label{Hmatter}
\eeq
where $R_\phi$ is the expansion 
scale factor when inflaton oscillations begin. At time $t_i$, 
when the saxion oscillations begin,
we set $m_{S_i} \simeq 3H(t_i)$ and find
\beq
\left(\frac{R_\phi}{R_i}\right)^3 = \frac{\lambda}{3 \sqrt{8}\pi}\, ,
\eeq
where $R_i$ is the scale factor when saxion oscillations begin.
The reheating completes at $R = R_R$, when
\beq
\left(\frac{R_\phi}{R_R}\right)^{3} =  \left(\frac{2 \Gamma_\phi}{3 H_I }  \right)^2 \, ,
\eeq
which leads to the following ratio of the scale factors
\beq
\label{ratioidphi}
\left(\frac{R_i}{R_R}\right)^{3} = \frac{8\sqrt{2} \pi}{3 \lambda} \left( \frac{\Gamma_\phi}{H_I} \right)^2\, .
\eeq

\subsection{Constraints for Kinetic Misalignment}
\label{subsec:KMM}

During inflation, quantum fluctuations of the axion and saxion fields are generated.
Because we consider axions as dark matter, the axion field fluctuations can be related to the power spectrum of the cold dark matter isocurvature perturbation~\cite{isoc}, which is given by
\beq
\label{isocurv}
\mathcal{P}_{S}(k_*) = \Bigg \langle \left(\frac{\delta \Omega_a}{\Omega_c}\right)^2 \Bigg \rangle = \Bigg \langle \left(\frac{\delta Y_{\theta}}{Y_{\theta}}\right)^2 \Bigg \rangle \, ,
\eeq
where we used Eq.~(\ref{ab1}) to relate $\Omega_a$ and $Y_\theta$, and assumed that axions make up all of the cold dark matter, so that $\Omega_a = \Omega_c$. Because the parameter $Y_{\theta}$ is a function of the angle $\theta$ and the saxion field $S$, the quantum fluctuation in $\delta Y_{\theta}$ is given by
\begin{equation}
\label{yfluc}
    \Bigg \langle \left(\frac{\delta Y_{\theta}}{Y_{\theta}}\right)^2 \Bigg \rangle = 
    \left(\frac{1}{Y_\theta} \frac{\partial Y_{\theta}}{\partial \theta_i}\right)^2 \langle\delta \theta_i^2 \rangle + 
    \left(\frac{1}{Y_\theta} \frac{\partial Y_{\theta}}{\partial S_i}\right)^2 \langle\delta S_i^2 \rangle  \, ,
\end{equation}
where $\sqrt{\langle \delta \theta_i^2 \rangle} = H_{I}/2 \pi S_i$ is the fluctuation of the initial angle and $\sqrt{\langle \delta S_i^2\rangle} = H_{I}/2 \pi$ is the fluctuation of the initial saxion field.

While the induced fluctuations from $\delta \theta_i$ can be tuned arbitrarily small if $\partial Y_\theta/\partial \theta_i \simeq 0$, those induced from $\delta S_i$ can not.
If the saxion oscillations begin before reheating, we can use Eq.~(\ref{yth2}) (see below) together with Eq.~(\ref{der8}) (see Appendix ~\ref{app:asym}), and we find that the power spectrum of dark matter isocurvature perturbations is given by
\begin{equation}
\label{ps1}
   \mathcal{P}_{S}(k_*) = \frac{(n - 3)^2}{\sqrt{2} \pi} \lambda \, ,
\end{equation}
where, for simplicity, we set the initial angle at the onset of angular rotations/radial oscillations to be $|\sin {n \theta_i}| = 1$.

Similarly, if the saxion oscillations begin after reheating, we use Eq.~(\ref{yth3}) (see below) to find the power spectrum of dark matter isocurvature perturbations
\begin{equation}
\label{ps2}
    \mathcal{P}_{S}(k_*) = \frac{\left(n - 5/2 \right)^2}{\sqrt{2} \pi} \lambda \, .
\end{equation}
From the CMB constraints we have~\cite{planck18}
\begin{equation}
    \beta_{\rm iso} \equiv \frac{P_S(k_*)}{P_{\zeta}(k_*) + P_S(k_*)} < 0.038 \quad (95\% \rm \, CL)\, ,
\end{equation}
where $P_S(k_*)$ and $P_{\zeta}(k_*) \simeq 2.2 \times 10^{-9}$ are the power spectra of isocurvature and adiabatic perturbations respectively with the pivot scale  $k_* = 0.05 \, \text{Mpc}^{-1}$. Therefore, we obtain
\beq
\label{powercons}
P_S(k_*) < 8.7 \times 10^{-11}\, .
\eeq
By combining the constraint~(\ref{powercons}) with Eqs.~(\ref{ps1}) and~(\ref{ps2}), we find the following upper bound for $\lambda$
\begin{equation}
    \lambda < \frac{4 \times 10^{-10}}{(n - x)^2} \, ,
    \label{isoa}
\end{equation}
with $x = 3$ if the saxion oscillations begin before reheating and $x = 5/2$ if the saxion oscillations begin after reheating. 

The saxion isocurvature fluctuations might also lead to a domain wall problem arising from power-law growth of modes with $k \simeq 0$. The origin of the growth is the following. In different Hubble patches, the radial and angular fields take slightly different values, due to isocurvature mode fluctuations, resulting in different initial rotation frequencies $\dot{\theta}_i$. As a result, the axion field values in different Hubble patches deviate from each other over time. This leads to domain wall formation once the QCD axion potential becomes non-negligible. As there is no point in space where the symmetry is restored, cosmic strings cannot form. Consequently, this is problematic even if the domain wall number is unity since in this case domain walls are not attached to cosmic strings. Although domain walls annihilate inside the horizon, this does not eliminate them from the entire Universe. The Universe would eventually become dominated by domain walls with sizes larger than the Hubble horizon. As shown in Appendix~\ref{app:dtheta}, this problem is avoided if quantum fluctuations generated during inflation are small enough,
\begin{equation}
n \frac{H_I}{2\pi} < f_a \times
\begin{cases}
1 & : \text{oscillation after reheating}  \\
\left( \frac{\sqrt{2}\pi^3}{15}g_* \right)^{-1/6}\left( \frac{\lambda^{1/2} H_I M_{\rm pl}}{T_R^2} \right)^{1/3} & : \text{oscillation before reheating}.
\end{cases}
\end{equation}
We find that this condition is violated in most of the allowed parameter space.

However, for $\epsilon \lesssim 0.8$ parametric resonance occurs and this domain wall problem is alleviated because the PQ symmetry is non-thermally restored, as discussed in Sec.~\ref{sec:PR}, and the formation of cosmic strings allows the domain walls to decay. In this case, axion dark matter can originate from either parametric resonance or kinetic misalignment. If the saxion is thermalized before $S$ reaches $f_a$, then the axion fluctuations are also thermalized due to the significant mixing of the two modes from the quartic coupling. Axions in the form of rotational oscillations persist due to the conservation of PQ charge and can contribute a dominant fraction to the dark matter density via kinetic misalignment.
The relative fraction from kinetic misalignment versus parametric resonance is discussed in the next section.

It is important to note another upper bound for $\lambda$. To ensure that the explicit PQ symmetry breaking effects do not significantly shift the CP conserving minimum of the axion, the additional contribution to the axion mass-squared from $V_A$, $m_{a,A}^2$, must be $\lesssim m_a^2 \, \theta_0$, where $m_a$ is given in Eq.~(\ref{mafa}) and $\theta_0 \simeq 10^{-10}$ is the current limit on CP violation determined from the experimental upper bound of the neutron electric dipole moment~\cite{NEDM_EX}.

From Eq.~(\ref{numden2}), we can relate $\epsilon$ to the coupling $A$ in $V_A$
as shown in Appendix \ref{app:asym} in Eq.~(\ref{der8}). If we solve for $A$, we find:
\begin{equation}
    A = \epsilon \, 2^{n/2 - 3} M_{\rm Pl}^{n-3} S_i^{4-n} \lambda^2 \csc{n \theta_i} \, .
\end{equation}
Using this expression in the potential, $V_A$, we find
\begin{equation}
    V_A = \frac{S^n S_i^{4-n} \epsilon \, \lambda^2 \,  \cos{n \theta}}{4n \sin{n \theta_i}}\, .
\end{equation}
Then the contribution to the axion mass from $V_A$ is
\begin{equation}
    \left| m_{a,A}^2 \right| = 2^{\frac{3n}{4}-5} n \pi ^{\frac{n}{2}-2} \epsilon \, f_a^{n-2} H_i^{4-n} \lambda ^{\frac{n}{2}}  \left| \csc \left(n \theta _i\right)  \right| \, ,
\end{equation}
where we have used Eq.~(\ref{si}) for the initial value for $S_i$
and have evaluated the mass at $S=f_a$ and $\theta = 0$.
If we set $\left| \csc \left(n \theta _i\right)  \right| = 1$, we find the following upper bound for $\lambda$
\begin{equation}
    \lambda < \frac{1}{2\sqrt{2} \pi} \left(\frac{H_I}{f_a} \right)^2 \left( \frac{32 \pi^2 \theta_0 f_a^2 m_a^2 }{n \epsilon H_I^4   } \right)^{2/n}\, .
\end{equation}

 This constraint is relaxed as $n$ is increased. By choosing $n = 10$, the explicit PQ breaking does not significantly affect the axion mass and does not further constrain the parameter space considered in this work.
 
If we compare the upper bounds~(\ref{isoa}) and the lower bound~(\ref{noinfl}), we find that the maximum viable value of $H_I$, that satisfies both constraints, is given by
\beq
H_{\rm{I, \, Max}} \simeq \frac{3 \times 10^{14}}{(n-x)} \, \text{GeV} \, ,
\eeq
where as before, $x = 3$ when oscillations begin before reheating and $x = 5/2$ otherwise. 

\begin{figure}[!ht]
\centering
\includegraphics[width=0.75\linewidth]{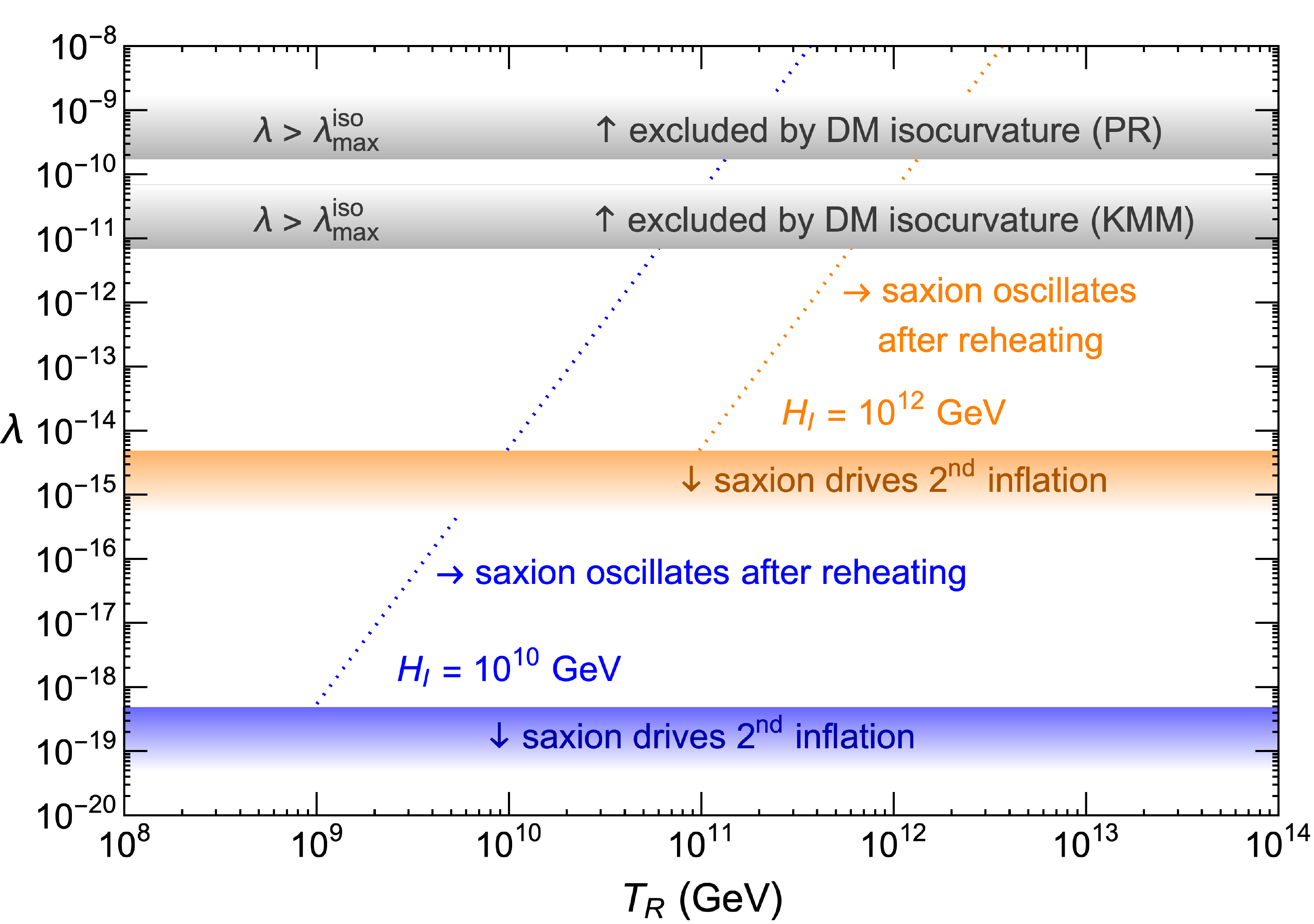}
\caption{The parameter space of the coupling $\lambda$ versus the reheating temperature $T_R$ for two values of $H_I$. To the left (right) of the sloped dotted line, saxion oscillations occur before (after) reheating. The gray regions exclude $ \lambda > \lambda_{\rm max}^{\rm iso}$~(\ref{isoa}) from dark matter isocurvature fluctuations for PR and KMM. Depending on the choice of $H_I$, the region below each colored boundary is excluded as the saxion potential drives a second period of inflation.}
\label{fig:radial}
\end{figure}

In Fig.~\ref{fig:radial}, we show the $\lambda$ vs $T_R$ plane. The upper limit on $\lambda$ from isocurvature fluctuations is shown by the gray region labeled KMM at large $\lambda$ for $n = 10$. Here, we consider two values of $H_I$ as labelled. For each value, the colored region shows the lower bound on $\lambda$ from Eq.~(\ref{noinfl}). Below these regions, a second period of inflation driven by the saxion field occurs. In the regions to the left (right) of the dotted sloped line, the saxion oscillations begin before (after) reheating. 

\subsection{Constraints for Parametric Resonance}
\label{subsec:PR}

For parametric resonance, the isocurvature constraint is modified due to the different dependence on $S_i$. The axion abundance is independent of $\epsilon$, and the power spectrum of dark matter isocurvature perturbations is
\begin{equation}
\label{psPR}
   \mathcal{P}_{S}(k_*) = \frac{(4-x)^2}{\sqrt{2} \pi} \lambda \, ,
\end{equation}
where $x = 3~(5/2)$ if the saxion oscillations begin before (after) reheating. This result can be understood by setting $n = 4$ in Eqs.~(\ref{ps1}) and (\ref{ps2}) because it is $V_0 \propto |P|^4$ rather than $V_{A} \propto P^n$ that determines the final dark matter abundance. The limit from isocurvature fluctuations is shown by the gray region labeled PR in Fig.~\ref{fig:radial}.

\section{Axion Dark Matter}
\label{sec:abund}

The axion abundance has two distinct origins: kinetic misalignment and parametric resonance. A key role is played by the parameter $\epsilon$, defined in Eq.~(\ref{numden2}), which determines the amount of initial PQ field rotation. The axion abundance from KMM is proportional to $\epsilon$ and once produced these axions survive, due to the conservation of PQ charge. PR occurs if $\epsilon$ is below the critical value of Eq.~(\ref{ep_crit}) and, as mentioned in Sec.~\ref{subsec:KMM}, axions produced from PR are depleted if thermalization occurs when $S \gg f_a$. In the following subsections, we analyze the parameter space for these two contributions separately. We discuss possible thermalization channels in Sec.~\ref{sec:therm}.

\subsection{Kinetic Misalignment}
\label{sec:KMM_DM}

In this subsection, we assume the axion abundance is produced entirely from KMM, which is the case when the PQ field is thermalized before $S$ relaxes to $f_a$. We first fix the value of $\lambda$ for various cases while taking into account the constraint from the isocurvature mode, and then discuss other constraints on the $(T_R,f_a)$ plane.

\begin{figure}
\centering
\includegraphics[width=0.65\linewidth]{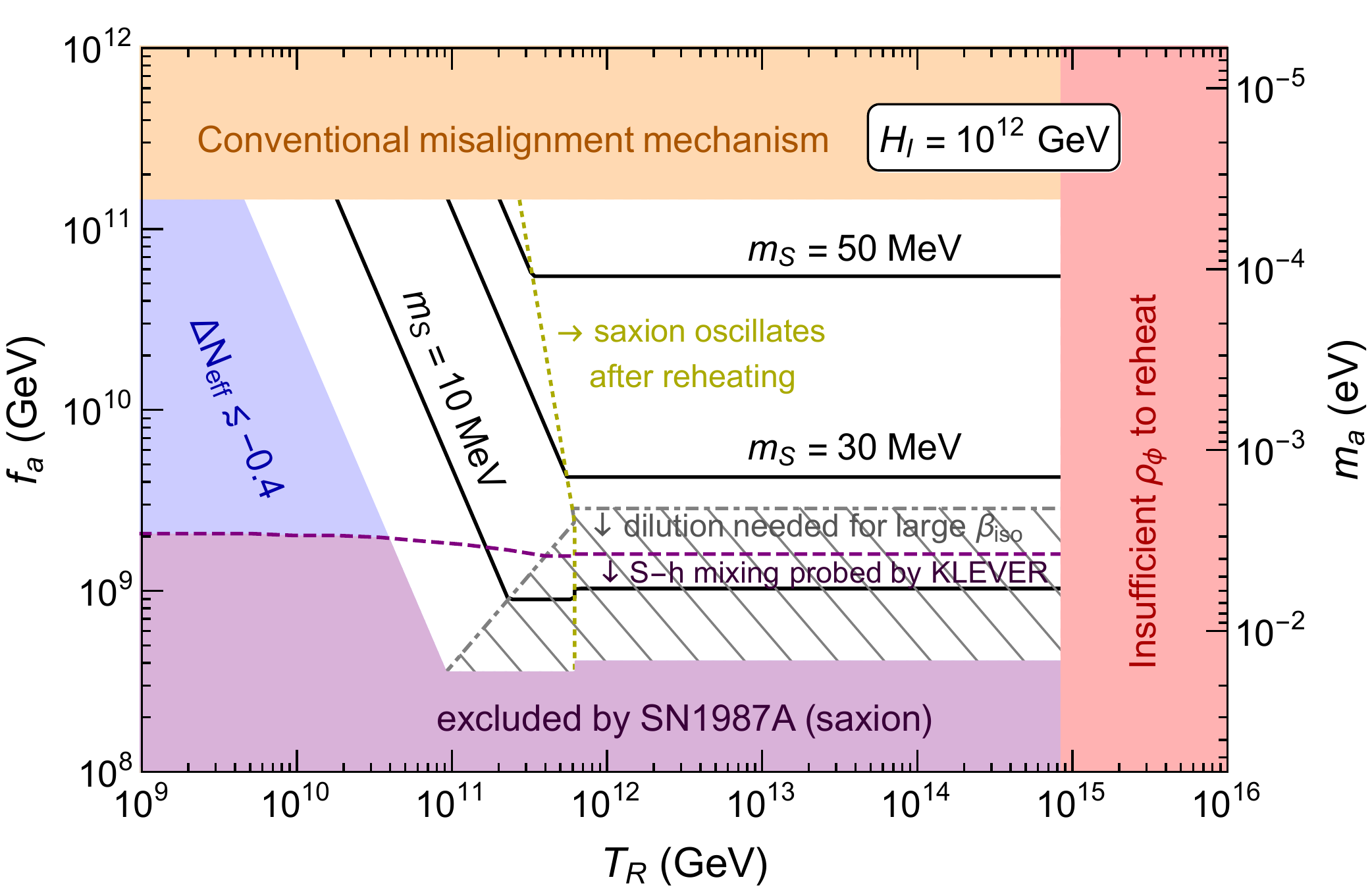}
\includegraphics[width=0.65\linewidth]{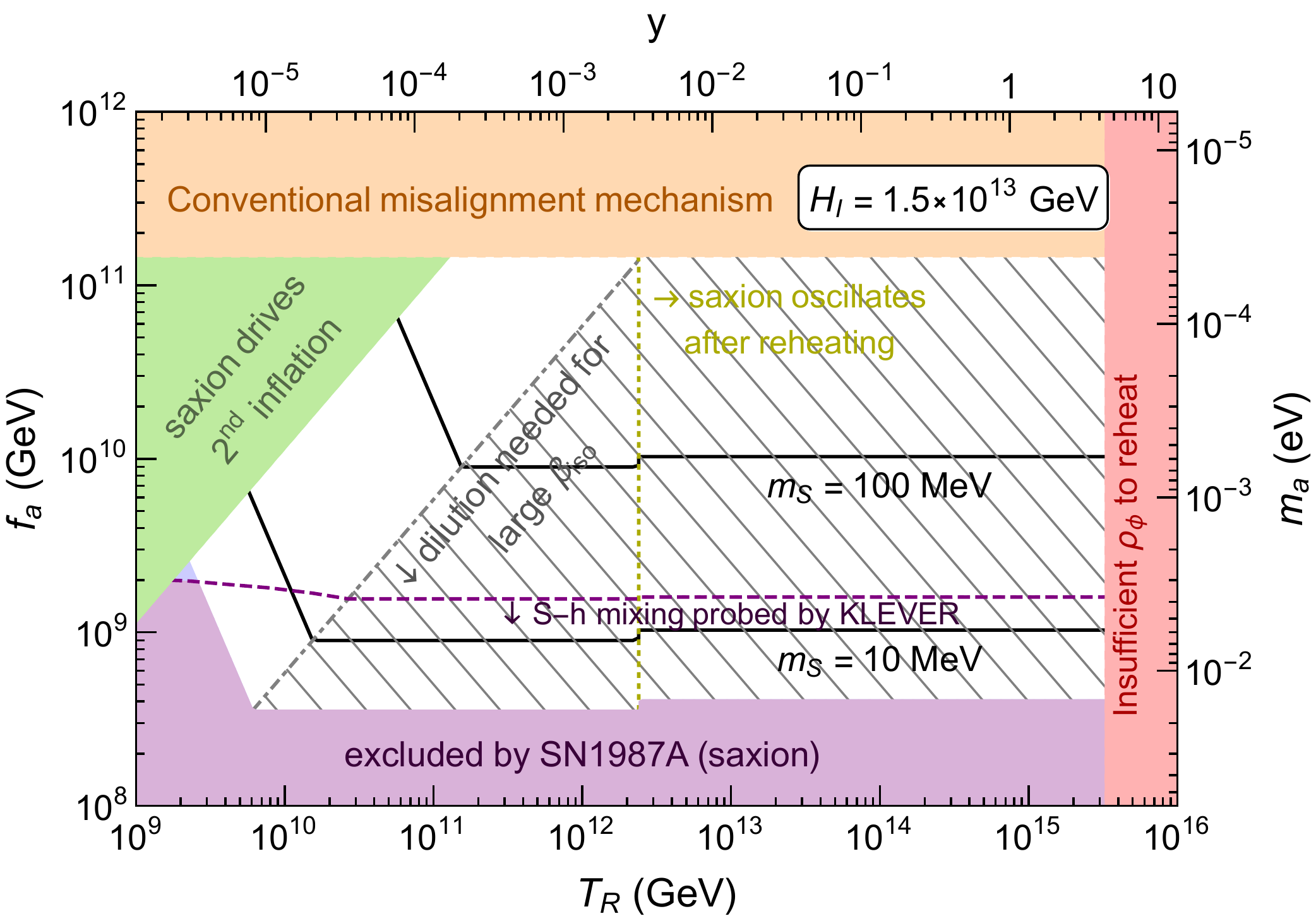}
\caption{Kinetic misalignment gives the observed dark matter abundance in the unshaded regions of the $(T_R, f_a)$ plane with black contours showing required values of $m_S = \sqrt{2} \lambda f_a$, for $\epsilon = 0.5$, $n=10$, and two values of $H_I$ as labelled.
Inside the gray hatched region, isocurvature modes are excessively produced, but can be diluted by entropy production.
To the right (left) of the yellow dotted line, radial oscillations begin after (before) the end of reheating. In the orange region, the conventional misalignment mechanism applies. 
The purple region shows the supernova constraint~\cite{astr2, newsn}. Below the purple dashed curve, the saxion-Higgs mixing required to evade the supernova constraint can be probed by searches for $K_L \rightarrow (\pi^0 + \rm{missing})$ by the KLEVER experiment~\cite{Ambrosino:2019qvz, Beacham:2019nyx}. The green region is excluded because the saxion causes a second inflationary era. 
In the blue area, the thermalized saxion decays will adversely affect BBN.
The red region  is unphysical because $T_R$ exceeds the maximum value allowed by $H_I$. In the lower panel, the value of the inflaton decay coupling $y$ is given on the upper axis using Eq.~(\ref{ycoup}) and $m_\phi = H_I/2$.}
\label{fig:fa_TR}
\end{figure}

Let us start with the case when saxion oscillations begin before reheating. After production, the initial PQ charge density of (\ref{numden2}) is diluted during the inflaton matter-dominated era, so that by the end of reheating it is
\begin{equation}
n_{\theta}(T_R) = n_{\theta_i} \left( \frac{R_i}{R_R} \right)^3 = n_{\theta_i} \left( \frac{H(T_R)}{H_i} \right)^2 \, .
\end{equation}
The axion yield at reheating is
\begin{equation}
Y_{\theta}= Y_\theta(T_R) = \frac{n_{\theta}(T_R)}{ s(T_R)} \simeq 6 \times 10^2 \, \epsilon \left( \frac{10^{-12}}{\lambda} \right)^{3/2} \left( \frac{H_I}{10^{12} \text{GeV}} \right) \left( \frac{T_R}{10^{11} \text{GeV}} \right) \, ,
\label{yth2}
\end{equation}
and we assume that the subsequent evolution is adiabatic so that $Y_\theta$ is constant.
We use Eq.~(\ref{ab1}) for the dark matter abundance with $Y_{\theta}$ from Eq.~(\ref{yth2}), and find
\begin{equation}
\label{ab2}
\Omega_a h^2 \simeq 0.2 \, \epsilon \left( \frac{10^{-12}}{\lambda} \right)^{3/2} \left( \frac{H_I}{10^{12} \, \text{GeV}} \right) \left( \frac{T_R}{10^{11} \, \text{GeV}} \right) \left( \frac{10^{10} \, \text{GeV}}{f_a}\right) \, \, .
\end{equation}
If we fix the dark matter abundance $\Omega_a h^2 \simeq 0.12$, we find the following expression for $\lambda$
\begin{align}
\label{cons1}
\lambda \simeq 10^{-12}  \, \epsilon^{2/3} \, \left( \frac{10^{10} \, \text{GeV}}{f_a} \right)^{2/3} \left( \frac{H_I}{10^{12} \, \text{GeV}} \right)^{2/3} \left( \frac{T_R}{10^{11} \, \text{GeV}} \right)^{2/3}.
\end{align}
We emphasize that these results are general and do not depend on any specific model of inflation.
In Fig.~\ref{fig:fa_TR}, we show contours of the saxion vacuum mass $m_S = \sqrt{2} \lambda f_a$ (solid black), which are determined using (\ref{cons1}) to fix $\lambda$ outside the gray hatched region, in the $f_a - T_R$ plane with $H_I = 10^{12}$ GeV (upper panel) and $H_I = 1.5 \times 10^{13}$ GeV (lower panel), assuming $\epsilon = 0.5$ and $n=10$.  Various constraints, including the gray hatched region, and prospects are explained below.

The axion abundance $\Omega_a h^2$ when saxion oscillations begin after reheating, is independent of the 
reheat temperature, $T_R$, as one would expect. 
In this case, we can calculate $Y_\theta$
directly at $T_i$ in Eq.~(\ref{ti}),
\begin{equation}
\label{yth3}
Y_{\theta} \simeq 2 \times 10^3 \, \epsilon \, \left( \frac{10^{-12}}{\lambda} \right)^{5/4} \left( \frac{H_I}{10^{12} \, \text{GeV}} \right)^{3/2} \, \, . 
\end{equation}
The dark matter abundance is given by
\begin{equation}
\label{ab3}
\Omega_a h^2 \simeq 0.8 \, \epsilon \left(\frac{10^{-12}}{\lambda} \right)^{5/4} \left(\frac{H_I}{10^{12} \, \text{GeV}} \right)^{3/2} \left( \frac{10^{10} \, \text{GeV}}{f_a}\right) \, \, .
\end{equation}
For $\Omega_a h^2 \simeq 0.12$, we find the following $\lambda$
\begin{align}
\lambda \simeq 4 \times 10^{-12}  \, \epsilon^{4/5} \, \left( \frac{10^{10} \, \text{GeV}}{f_a} \right)^{4/5} \left( \frac{H_I}{10^{12} \, \text{GeV}} \right)^{6/5},
\label{oh2lim2}
\end{align}
which is independent of reheating temperature $T_R$. Therefore, the contours of constant $m_S$ in Fig.~\ref{fig:fa_TR} are horizontal. 
When $T_i = T_R$, the expressions for dark matter abundance $\Omega_a h^2$ in Eqs.~(\ref{ab2}) and (\ref{ab3}) are identical. 

Inside the gray hatched region, the value of $\lambda$ needed to obtain the observed dark matter density, shown in Eqs.~(\ref{cons1}) and (\ref{oh2lim2}), exceeds the bounds found in Eq.~(\ref{isoa}) from isocurvature fluctuations. The bound can  be evaded by taking smaller $\lambda$. Axions are overproduced by KMM, but may be diluted by entropy production. (Such dilution may be provided by the saxion, although we do not investigate such a possibility in detail.) To maximize the allowed parameter space, inside the gray hatched region, we take a constant $\lambda$ saturating the isocurvature constraint, assuming an appropriate amount of entropy production. As a result the prediction on $m_S$ is independent of $T_R$, except for the step because of the change of the isocurvature constraint at $T_i = T_R$.

The yellow dotted line shown in Fig.~\ref{fig:fa_TR}, defined by using Eq.~(\ref{boundary}),
corresponds to the boundary where saxion oscillations begin at the completion of reheating. Outside the gray hatched region, using Eqs.~(\ref{boundary}) and (\ref{cons1}), the yellow dotted line is given by
\begin{equation}
\label{fa2}
f_a =  4 \times 10^{13} \, \text{GeV} \, \epsilon \left( \frac{H_I}{10^{12} \, \text{GeV}} \right)^4 \left( \frac{10^{11} \, \text{GeV}}{T_R} \right)^5 \, ,
\end{equation}
while it is vertical inside the gray hatched region since $\lambda $ is constant there.
To the left of the yellow dotted line, oscillations begin before the end of reheating. To the right of the yellow dotted line, oscillations begin in the radiation-dominated era after reheating.

We display various constraints on the parameter space in Fig.~\ref{fig:fa_TR}. In the orange region, the conventional misalignment mechanism is applicable, as is shown in Eq.~(\ref{eq:fa_KMM}). The green region violates Eq.~(\ref{noinfl}) and results in a second period of inflation by the saxion, and axions are under-produced. The red region is unphysical since the radiation energy density at the completion of reheating exceeds the energy density during inflation.

We additionally impose the supernova constraint, which requires $f_a \gtrsim 10^{8}-2\times 10^9 \, \text{GeV}$~\cite{astr2,newsn}, so that energy loss due to axion or saxion emission does not conflict with the neutrino observations from SN1987A\footnote{A similar limit is available from the cooling of the neutron star in Cas A \cite{casA}.}. This bound can be evaded for $m_S > 4$ MeV; below the purple dashed curve, the saxion can be trapped inside the core and thus evade the supernova cooling constraint by a mixing with the Higgs, whose required magnitude is consistent with the couplings considered below. The required magnitude of the mixing can be probed by KLEVER~\cite{Ambrosino:2019qvz, Beacham:2019nyx} through the decay $K_L \rightarrow \pi^0 +S$, with the latter being invisible.
As discussed in Sec.~\ref{sec:decay}, $m_S < 4$ MeV is excluded due to the effect of saxions on the CMB.  In Sec.~\ref{sec:decay} we also show that if cosmological saxions are depleted by an interaction $S^2 H^\dagger H$, then much of the allowed region of parameter space can be probed by KLEVER.

Consider the lower panel of Fig.~\ref{fig:fa_TR} with $H_I = 1.5 \times 10^{13}$ GeV. Given our constraints on $\lambda$ in Eqs.~(\ref{noinfl}) and (\ref{isoa}), we see that without entropy production after the inflaton decays, we must have $10^{12}~{\rm GeV} \gtrsim T_R \gtrsim 10^9$ GeV and $f_a \gtrsim 4 \times 10^8$ GeV from the supernova bound. Entropy production allows higher $T_R$.

\subsection{Parametric Resonance}
\label{sec:PR_DM}
 
 While the discussion in Sec.~\ref{sec:KMM_DM} is dedicated to KMM, the same formulae can be used to obtain the contribution from PR, which is present for $\epsilon \lesssim 0.8$ according to Eq.~(\ref{ep_crit}). The dark matter abundance from PR is given by
\begin{equation}
\label{rho_PR}
\frac{\rho_a}{s} = m_a Y_a = m_a \frac{n_a}{s} ,
\end{equation}
with the initial number density $n_{a_i}$ given in Eq.~(\ref{axion_PR}). Comparing Eqs.~(\ref{rho_KMM}) and (\ref{rho_PR}) as well as Eqs.~(\ref{numden2}) and (\ref{axion_PR}), we observe that, by effectively setting $\epsilon = 1/2$, the formulae in Sec.~\ref{sec:KMM_DM} can be directly translated into those for PR even when $\epsilon \ll 1$. In this subsection we will assume $\epsilon \ll 1$, which renders the KMM axion abundance negligible, and we also assume that
the axions produced by PR are not thermalized, so that the final axion abundance is produced entirely from PR. 

\begin{figure}
\centering
\includegraphics[width=0.65\linewidth]{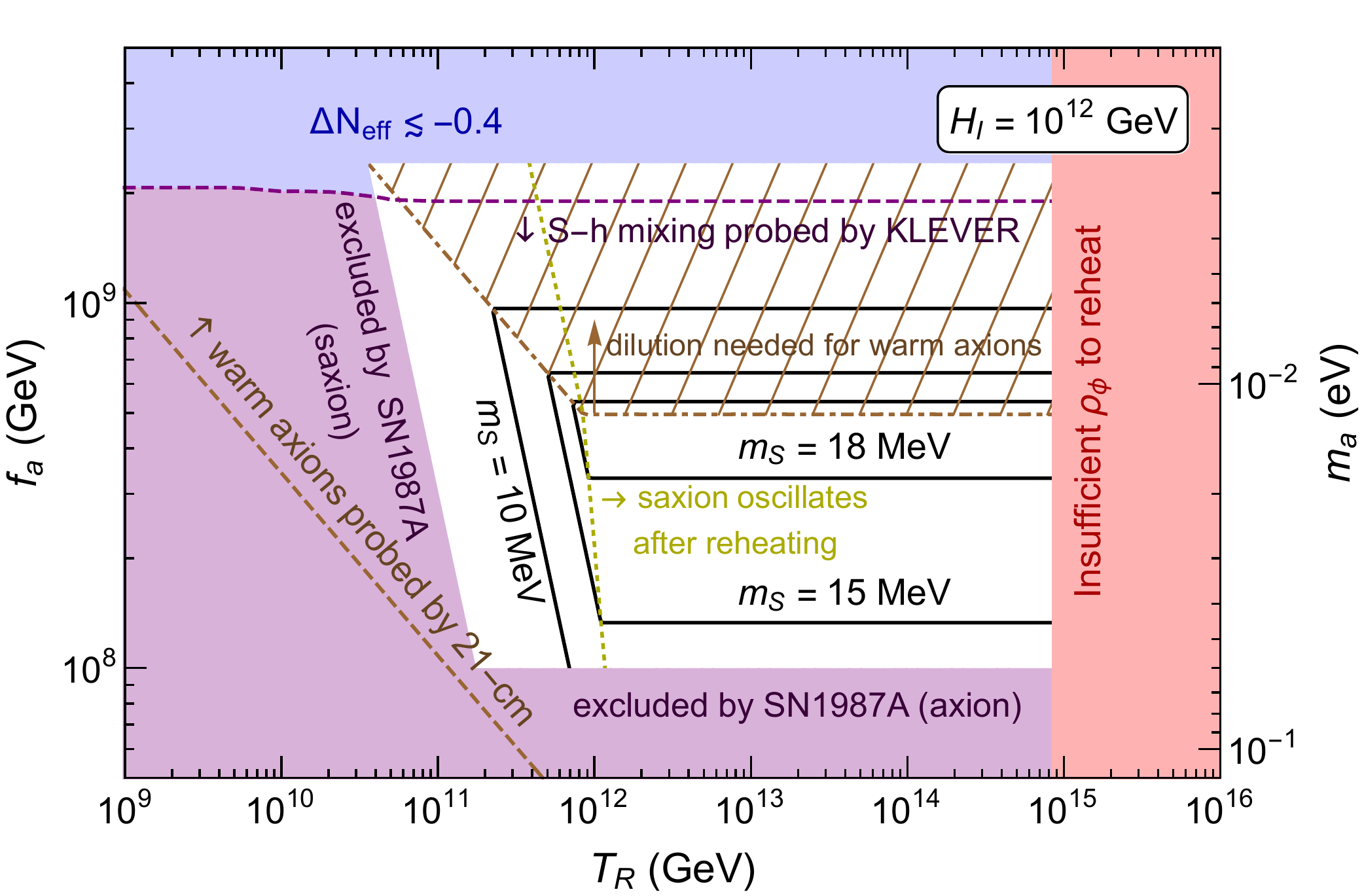}
\includegraphics[width=0.65\linewidth]{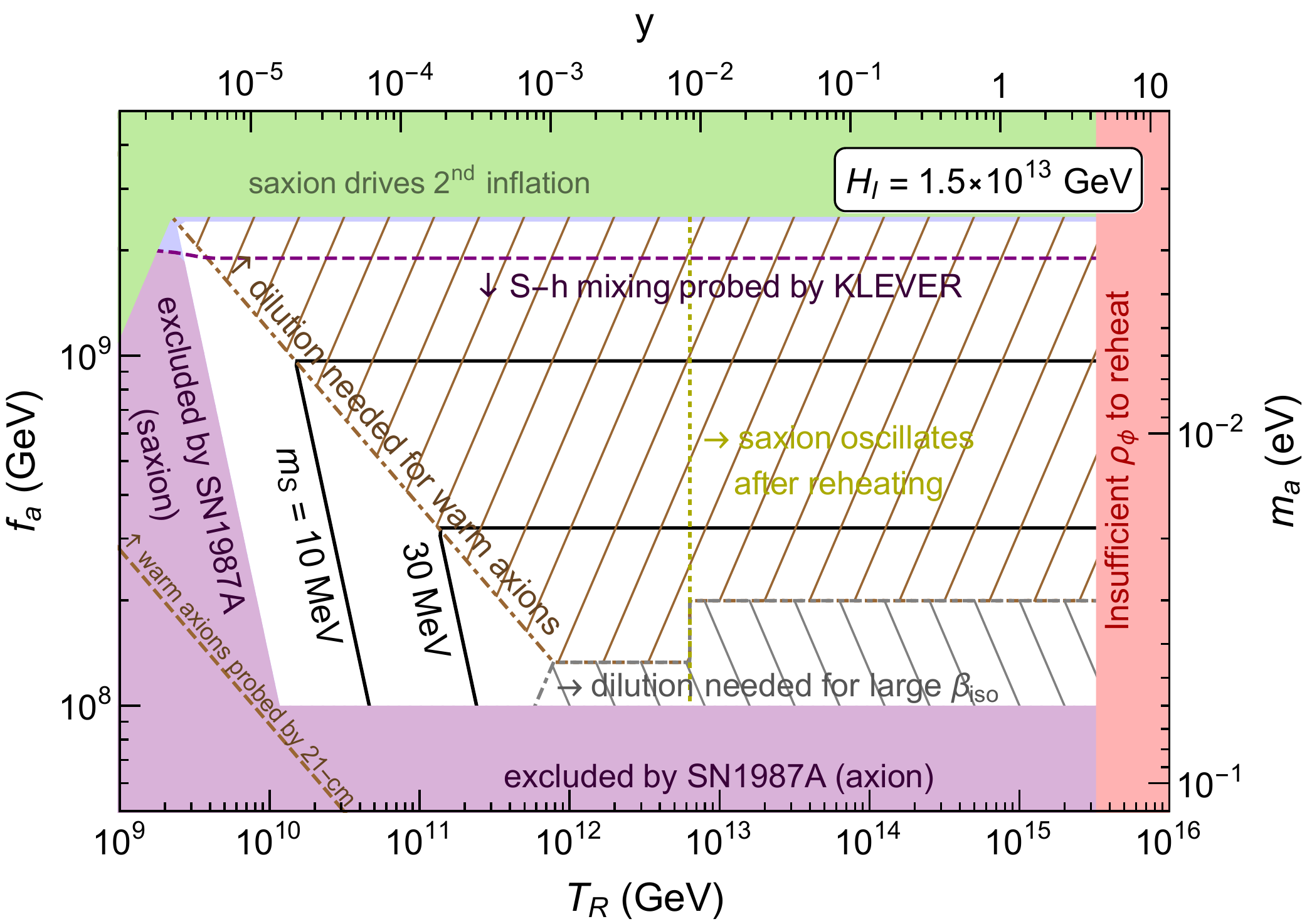}
\caption{Same as Fig.~\ref{fig:fa_TR} but for $\epsilon \ll 1$ so parametric resonance is the dominant source of axion dark matter.
Inside the brown hatched region, axion dark matter is too warm unless entropy production occurs. Above the brown dashed line, the warmness of axion dark matter has detectable effects on cosmic 21-cm lines. In the unshaded region below the purple dashed curve, the saxion-Higgs mixing required to evade the supernova constraint can be probed by KLEVER.}
\label{fig:fa_TR_PR}
\end{figure}

Axions produced by PR may be warm enough to affect structure formation.
The warm dark matter constraint can be phrased in terms of the velocity of dark matter at the temperature of 1 eV, $v_{\rm DM}({\rm eV})$. The current bound is $v_{\rm DM}({\rm eV}) \lesssim 10^{-4}$~\cite{Viel:2013apy}, while cosmic 21-cm signals can probe $v_{\rm DM}({\rm eV}) \gtrsim 10^{-5}$~\cite{Sitwell:2013fpa}. From the initial axion momentum $k_a \simeq m_{S_i}$ at the onset of the saxion oscillations, we can derive the velocity of an axion at $T = 1$ eV using redshift invariant quantities such as $k_a^3 / \rho_\phi$ during the matter-dominated era and $k_a^3 / s$ during the radiation-dominated era. For the case when the saxion oscillates before the end of reheating, the axion velocity as a function of temperature after reheating is
\begin{equation}
\label{eq:va_1}
v_a(T) \simeq 7 \times 10^{-5} \left( \frac{T}{1~{\rm eV}} \right) \left( \frac{f_a}{10^9~{\rm GeV}} \right)^{8/9} \left( \frac{H_I}{10^{12}~{\rm GeV}} \right)^{4/9}  \left( \frac{T_R}{10^{10}~{\rm GeV}} \right)^{4/9} .
\end{equation}
On the other hand, when saxion oscillations begin after the end of reheating
\begin{equation}
\label{eq:va_2}
v_a(T) \simeq 7 \times 10^{-5} \left( \frac{T}{1~{\rm eV}} \right) \left( \frac{f_a}{3 \times 10^8~{\rm GeV}} \right)^{4/5} \left( \frac{H_I}{10^{12}~{\rm GeV}} \right)^{4/5} .
\end{equation}

We fix the value of $\lambda$ as we did in Sec.~\ref{sec:KMM_DM}. Contours of $m_S$ are shown by black lines in Fig.~\ref{fig:fa_TR_PR}. We take into account the isocurvature constraint (the gray hatched region), as well as the warmness constraint (the brown hatched region). The isocurvature constraint is greatly relaxed compared to the case with KMM in Fig.~\ref{fig:fa_TR} due to a milder dependence on $S_i$ as explained below Eq.~(\ref{psPR}). The warmness constraint is more severe for high $m_S$ because the initial momentum is larger and for high $f_a$ because the axion vacuum mass is smaller. 

The warmness constraint can be avoided by taking smaller $\lambda$, overproducing axions by the PR, and diluting them by entropy production. Inside the brown hatched region, we take the largest possible $\lambda$ while satisfying the warmness constraint
\begin{equation}
\lambda \lesssim 7 \times 10^{-12} \left( \frac{10^9~{\rm GeV}}{f_a} \right)^2 .
\end{equation}
The prediction of $\lambda$, and hence that of $m_S$, become independent of $T_R$. We note that the blue, green, and purple regions to the left of the brown hatched region are identical to those outside the gray hatched region in Fig.~\ref{fig:fa_TR}. Inside the brown/gray hatched regions, they appear differently because brown and gray hatched regions, where constraints become $T_R$-independent, cover different areas between the two figures.

In Fig.~\ref{fig:fa_TR_PR}, the viable parameter space is shown for axion dark matter dominantly produced from PR. According to Eqs.~(\ref{eq:va_1}) and (\ref{eq:va_2}), the PR produced axions are warm enough to leave imprints on 21-cm lines above the brown dashed line. The scenario of axion dark matter from PR, applicable when $\epsilon \ll 1$, makes sharp predictions for $f_a$, $m_S$, the saxion-Higgs mixing, and warmness of dark matter. Remarkably, the entire allowed parameter space can be probed by both KLEVER and 21-cm lines.

 \section{Thermalization of Saxion Oscillations}
 \label{sec:therm}
 At later times, unless the saxion oscillations and fluctuations thermalize or decay, they will eventually come to dominate the energy density of the Universe. In this section, we show that the saxion can successfully thermalize before the domination occurs.
 
 Scattering with gluons occurs at a rate~\cite{Bodeker:2006ij, Laine:2010cq, Mukaida:2012qn}
 \beq
 \label{eq:Gamma_gluons}
\Gamma_{\rm th} \, \simeq \, b \frac{T^3}{f_{\rm eff}^2},
 \eeq
 where $f_{\rm eff} \simeq \max(S,f_a)$ and $b\simeq \alpha_3(T)^2/100 \simeq 10^{-5}$. Here $\alpha_3(T)$ is the strong gauge coupling constant evaluated at temperature $T$. We find that saxion oscillations thermalize only in a small portion of the parameter space, with low values of $f_a$, which is already excluded by other constraints for both values of $H_I$ shown
in Fig.~\ref{fig:fa_TR}. Thus, thermalization does not occur in any of the allowed regions of parameter space by scattering with gluons.
 
Another possible thermalization channel of the radial mode is via scattering with light PQ charged fermions $\psi$ and $\bar{\psi}$, which may also have a color charge. Such a Yukawa interaction $z P \psi \bar{\psi}$ leads to a scattering rate~\cite{Mukaida:2012qn}
\beq
\Gamma_{\rm th} \, \simeq \, 0.1 \, z^2 T,
\eeq
where $\psi$ and $\bar{\psi}$ are assumed to be in thermal equilibrium, requiring that $m_\psi (T) \simeq z f_{\rm eff} < T$. Therefore, with the largest possible $z$ at a given $T$, this scattering rate follows the same functional form as Eq.~(\ref{eq:Gamma_gluons}) but with $b \simeq 0.1$.
In much of the parameter space with $f_a \lesssim 10^{10}-10^{11}$~GeV, such a scattering with maximal $z$ would lead to the thermalization of the saxion oscillations before $S$ relaxes to $f_a$, while thermalization after $S$ relaxes to $f_a$ also occurs for lower values of $z$. In the former case, the axion fluctuations are also thermalized because the quartic coupling efficiently mixes saxions and axions. Thus, the survival of the PR axions discussed in Sec.~\ref{sec:PR_DM} depends on the existence and coupling strength of such light quarks. If thermalization occurs before $S$ relaxes to $f_a$, then the final axion abundance originates predominantly from KMM discussed in Sec.~\ref{sec:KMM_DM} even if both contributions are initially present with $\epsilon \lesssim 0.8$ based on~Eq.~(\ref{ep_crit}). In summary, we find that both scenarios studied in Sec.~\ref{sec:abund} are possible and may contribute to the present axion density.

A potentially more effective thermalization process is possible
if there is a coupling between the PQ field
and the Standard Model Higgs, giving a coupling $\xi^2 \, S^2 H^\dagger H$. While scatterings $S\, H \to S\, H$ are not effective at thermalization (we must have
$f_a \lesssim \sqrt{\lambda} \xi^4 M_{\rm Pl}^3/H_I T_R$), $S\, H \to H\, Z$ may be effective.  The rate for such scatterings is given by
\beq
\label{thhiggs}
\Gamma_{\rm th} =\alpha_2(T) \frac{ \xi^4 f_{\rm eff}^2}{T} \, ,
\eeq
where $\alpha_2(T) = g_2(T)^2/4 \pi$, and we have taken the VEV $\langle S \rangle = f_a$ on one of the external saxions. We assume that the thermalization process occurs at high temperatures, and we will use $\alpha_2(T) = \alpha_2 \simeq 1/30$. In writing (\ref{thhiggs}), we have implicitly assumed that $\xi f_a < T$. If $\xi f_a \gg 125$ GeV, we are required to tune the contribution to the Higgs mass from $\xi^2 \, S^2 H^\dagger H$ with an $S$-independent contribution. 

We first consider thermalization when $R_i < R_R$, that is when saxion oscillations begin before reheating ends.
For $S \gg f_a$, the saxion energy density
 is dominated by the quartic term and falls as $\rho_S \sim R^{-4}$. Below $T_R$,
 we can then assume that $\rho_S \sim T^4$, and that the amplitude of the oscillations scales as $S \sim T$. 
  At reheating, using Eqs.~(\ref{si}) and~(\ref{ratioidphi}), we can write the amplitude of $S$ as
\beq
S_R = S_i \left( \frac{R_i}{R_R} \right) = \frac{2^{5/12}}{3^{1/3}\pi^{1/6} \lambda^{5/6}}H_I^{1/3}\Gamma_\phi^{2/3} \, .
 \eeq
 For $S < S_R$, we can write $S = S_R T/T_R$.
 
 We define $T_S$ as the temperature when the potential $V_0$ is no longer dominated by the quartic term and is dominated by the quadratic term instead, which occurs when $S = \sqrt{2} f_a$, then $T_S  = \sqrt{2} f_a T_R / S_R$.
 The temperature at which thermalization occurs is determined by $\Gamma_{\rm th} \simeq 3H$. 
Since $f_{\rm eff}$ is temperature dependent when $S \gg f_a$,    
the expression for the thermalization temperature depends on whether $T_{\rm th}$ is greater or less than $T_S$.  While thermalization may be possible when $T > T_S$, the evolution of $S$ and $H$ are complicated by the rapidly changing masses of $S$ and $H$, and a full treatment of this case is beyond the scope of the present work. 

We can, however, address the case when $T_{\rm th} < T_S$. We estimate $T_{\rm th}$ by setting the scattering rate~(\ref{thhiggs}) equal to $3H$ and find
\begin{eqnarray}
T_{\rm th} & = & \alpha_2^{1/3} \, \left(\frac{10}{\pi^2 g_*} \right)^{1/6} \xi^{4/3} f_{\rm eff}^{2/3} M_{\rm Pl}^{1/3}, \quad T < T_S .
\label{th4}
\end{eqnarray}
To ensure that $T_{\rm th} < T_S$, we have an upper bound on $\xi$
\begin{equation}
\label{facrit}
    \xi < 2 \times 10^{-6} \, \epsilon_{\rm eff}^{5/12} \left(\frac{H_I}{10^{12} \, \text{GeV}} \right)^{1/6} \left(\frac{T_R}{10^{11} \, \text{GeV}} \right)^{1/6} \left(\frac{10^{10} \, \text{GeV}}{f_a} \right)^{1/6} \, ,   
\end{equation}
which is easy to satisfy across the plane shown in Fig.~\ref{fig:fa_TR}.
In this section, the $\epsilon$ dependence appears because $\lambda$ is fixed to reproduce the dark matter abundance. For this reason, we define $\epsilon_{\rm eff} = \epsilon$ when the final axion abundance is dominantly produced from KMM, and $\epsilon_{\rm eff} = 1/2$ when it is instead from PR as discussed below Eq.~(\ref{rho_PR}).

Note that $\xi$ must not be so small that thermalization takes place after saxions come to dominate the total energy density. To calculate the temperature when the saxion begins dominating the energy density, $T_M$, we first calculate the redshift invariant quantity, i.e., the saxion yield $Y_S = n_S/s$ for $T_i < T_R$. We compare the saxion energy density $m_S Y_S$ to the thermal bath and find that the saxion starts dominating the energy density when
\beq
T_{\rm M} = \frac{H_I f_a T_R}{ 2^{9/4} \sqrt{3 \pi } \sqrt{\lambda} M_{\rm Pl}^2} \, .
\label{TM}
\eeq
The requirement that $T_{\rm th} > T_{\rm M}$ leads to the constraint on $\xi$,
\beq
\label{xicond1}
\xi > 6 \times 10^{-9} \, \epsilon_{\rm eff}^{-1/4} \left(\frac{H_I}{10^{12} \, \text{GeV}} \right)^{1/2} \left(\frac{T_R}{10^{11} \, \text{GeV}} \right)^{1/2} \left(\frac{f_a}{10^{10} \, \text{GeV}} \right)^{1/2} \, ,
\eeq
where we use $f_{\rm eff} \simeq f_a$.
So long as $S(T_{\rm th})/f_a \sim (T_{\rm th}/T_S)^{3/2}$, this naive treatment of thermalization may be valid, though as noted above, a full treatment of the dynamics of $S$ and $H$ is warranted and left for future work.

When radial oscillations begin after reheating and $T_i < T_R$, $T_S = \sqrt{2} f_a T_i/S_i$, where $S_i$ and $T_i$ are given in 
Eqs.~(\ref{si}) and (\ref{ti}) respectively. 
The condition (\ref{facrit}) for $T_{\rm th} < T_S$ is independent of $T_R$, and is given by 
\beq
\xi < 3 \times 10^{-6} \, \epsilon_{\rm eff}^{9/20}  \left(\frac{H_I}{10^{12} \, \text{GeV}} \right)^{3/10} \left( \frac{10^{10} \, \text{GeV}}{f_a}\right)^{1/5} \, ,
\label{facrit2}
\eeq
and is again easily satisfied. To find $T_M$ for $T_i < T_R$, the calculation is similar to Eq.~(\ref{TM}) but the ratio of the saxion yield to that of the inflaton $Y_S/Y_\phi$ is a constant until reheating ends and $Y_S$ is invariant afterwards. In this case, saxion domination occurs at
\beq
T_{\rm M} = \frac{ 5^{1/4} H_I^{3/2} f_a}{2^{19/8} 3^{1/4} \pi ^{5/4} g_*^{1/4} \lambda^{1/4}
    M_{\rm Pl}^{3/2}}  \, .
    \label{TM2}
\eeq
The condition $T_{\rm th} > T_{\rm M}$ is now given by
\begin{equation}
\label{xicond2}
\xi > 10^{-8} \, \epsilon_{\rm eff}^{-3/20} \left(\frac{H_I}{10^{12} \, \text{GeV}} \right)^{9/10} \left(\frac{f_a}{10^{10} \, \text{GeV}} \right)^{2/5}  \, .
\end{equation}

To ensure that Higgs scattering is effective at thermalization, Higgs bosons must be present in the thermal bath at $T_{\rm th}$, requiring $m_H < T_{\rm th}$. 
If we expand the saxion field $S$ about its vacuum value, $S = \tilde S + f_a$, we can write
\beq
m_H^2 = -m^2 + \xi^2 f_a^2 + 2 \xi^2 f_a \tilde S,
\label{mH}
\eeq
where we have dropped the contribution from $\tilde S^2$.  In (\ref{mH}) we assume a cancellation between the first two terms
so that their sum returns the experimentally determined Higgs mass of 125 GeV when $\tilde S = 0$. The third term, however, is oscillatory
with frequency $m_S$ and at $T < T_S$ has an amplitude which falls as $\tilde S\sim T^{3/2}$, and $\tilde S$ can be calculated by setting the yield $Y_S = n_S/s$ equal to the yield $Y_S(T_{\rm th})$ when the thermalization occurs. If $T_{\rm th} < T_S$, the constraint $m_H^2 < T_{\rm th}^2$, where $m_H^2 = 2 \xi^2 f_a \tilde S$, becomes
\beq
\xi < 4 \times 10^{-6} \, \epsilon_{\rm eff}^{5/8} \left(\frac{H_I}{10^{12} \, \text{GeV}} \right)^{1/4} \left(\frac{T_R}{10^{11} \, \text{GeV}} \right)^{1/4} \left(\frac{10^{10} \, \text{GeV}}{f_a} \right)^{3/4} \, ,
\label{mhtth1}
\eeq
for $T_i > T_R$. Similarly, when $T_i < T_R$, we find
\beq
\xi < 5 \times 10^{-6} \, \epsilon_{\rm eff}^{27/40}  \left(\frac{H_I}{10^{12} \, \text{GeV}} \right)^{9/20} \left( \frac{10^{10} \, \text{GeV}}{f_a}\right)^{4/5} \, .
\label{mhtth2}
\eeq
For our nominal value of $H_I = 1.5 \times 10^{13} \, \text{GeV}$, we find that
$\xi \simeq 5 \times 10^{-7}$ simultaneously satisfies the constraints derived in this section across the entire unshaded $T_R - f_a$ region shown in Fig.~\ref{fig:fa_TR} and thus thermalization is efficient. For $H_I = 10^{12} \, \text{GeV}$, $\xi \simeq 10^{-7}$ allows the entire viable parameter space. We note however that there is a further constraint related to $\xi$ from successful depletion of the thermalized saxions, which we discuss in the next section.

Before concluding this section, we note that
during a fraction of an oscillation period, the Higgs mass becomes light enough to allow
for saxion decay to two Higgs bosons. 
Depending on the choice of parameters ($f_a, T_R, H_I, \xi$), decay may be still more efficient than scattering in the dissipation of the saxion condensate. 
As scatterings can be made efficient everywhere in the parameter space of interest, we leave the more detailed interplay between scattering and decay for future work. 

\section{Decay of Thermalized Saxions}
\label{sec:decay}

The final constraint of importance is the decay of the thermalized saxions before Big Bang Nucleosynthesis (BBN). The saxion mass is $m_{S} = \sqrt{2} \lambda f_a$ and could range from 0.1~MeV to 1 GeV for $\lambda$ between $10^{-12} - 10^{-11}$ and $f_a$ between $10^8 - 10^{11}$ GeV. If present
at late times, saxions could begin to dominate the energy density leading to a period of late matter domination or late decay and potentially spoiling the successful predictions of BBN~\cite{foyy}. 

In the absence of the coupling $\xi^2 \, SSH^\dagger H$,
the dominant channel for $S$ decay is to two light axions \cite{chh}. However, the resulting axions would contribute to the number of light degrees of freedom
and using the BBN bound on $N_{\rm eff}$, 
$\Delta N_{\rm eff} < 0.17$ \cite{foyy}, all of the 
parameter space previously discussed would be excluded.

The coupling of saxions to $HH$ induces a decay channel for $S$ to light particles such as $\mu^+ \mu^-$ or $e^+ e^-$ through the mixing of $S$ and $H$. The decay rate is given by
\beq
\Gamma_S = \frac{1}{8\pi} \theta_{SH}^2 y_f^2 m_S \, , 
\eeq
where $y_f$ is the Yukawa coupling of the final state to the Higgs boson. The mixing angle is 
\beq
\theta_{SH}  \simeq 2 \sqrt{2} \xi^2 \frac{f_a v}{m_h^2} \, ,
\eeq
where $v = 174$ GeV is the Higgs VEV. The condition for decay
in the radiation-dominated epoch is $\Gamma \gtrsim 2 H$ and the decay temperature is given by
\beq
T_{SD}^2 = \frac{\theta_{SH}^2 y_f^2}{16 \pi}\sqrt{\frac{90}{\pi^2 g_*}} M_{\rm Pl} m_S \, ,
\label{eq:TSD}
\eeq
where now $g_* = 10.75$.
BBN will be unaffected as long as $T_{SD} \gtrsim 2$ MeV.

However, $T_{SD} \gtrsim 2$ MeV is not a sufficient condition for problem-free decay. For $T_{SD} \sim m_S$, saxions are kept in equilibrium by decays and inverse decays, and saxion decay
continues past $T_{SD}$, injecting
radiation into the $e^+ e^- \gamma$
background. If significant injection occurs after neutrinos are decoupled around $T=2$ MeV, neutrinos become relatively cooler,
leading to a decrease in the effective number of neutrinos, $N_{\rm eff}$. Using the lower bound \cite{foyy}, $\Delta N_{\rm eff} > -0.44$,
leads to a lower bound on $m_S \gtrsim 4$ MeV to ensure that the saxion abundance is sufficiently Boltzmann suppressed when neutrinos decouple. 

The limits imposed by late saxion decay are also shown in Figs.~\ref{fig:fa_TR} and \ref{fig:fa_TR_PR}. Regions in which 
$m_S \lesssim 4$ MeV are shaded blue, while the constraint $T_{SD}> 2$ MeV is satisfied in the viable parameter space by choosing the largest value of $\xi$
allowed by Eqs.~(\ref{facrit}), (\ref{facrit2}), (\ref{mhtth1}), and (\ref{mhtth2}). When $H_I = 10^{12}$ GeV, this constraint is quite important and excludes much of the parameter space with lower values of $T_R$ ($\lesssim 10^{10}$ GeV) in Fig.~\ref{fig:fa_TR} and also higher values of $f_a$ ($\gtrsim 2 \times 10^9$ GeV) in Fig.~\ref{fig:fa_TR_PR}. 
In contrast, when $H_I = 1.5 \times 10^{13}$ GeV, the constraint only
excludes a small corner of previously allowed space at $f_a \simeq 2 \times 10^9$ GeV, again stemming from the lower mass limit on $m_S$ of 4 MeV.

Using (\ref{eq:TSD}), the requirement $T_{SD} > 2$ MeV leads to a lower bound on $\theta_{SH}$
\beq
\theta_{SH}^2 \, > \, 2 \times 10^{-9} \left( \frac{4 \, \rm{MeV}}{m_S} \right).
\label{eq:KL}
\eeq
The KLEVER experiment~\cite{Ambrosino:2019qvz, Beacham:2019nyx}, designed to measure the branching ratio for $K_L \rightarrow \pi^0 \bar{\nu} \nu$, could also see a signal for $K_L \rightarrow \pi^0 S$, reaching a sensitivity to $\theta_{SH}^2$ of $3 \times 10^{-10}$. Hence, if the $S^2 H^\dagger H$ operator is the origin of saxion thermalization and decay, KLEVER will see a signal over much of the allowed parameter space of Fig.~\ref{fig:fa_TR}. 

\section{Discussion}
\label{sec:disc}

While there are many potential dark matter candidates under both theoretical and experimental investigation, very few of them are motivated by extensions of the Standard Model built to resolve deficiencies in the Standard Model. Two classic examples are the lightest supersymmetric particle in supersymmetric extensions of the Standard Model \cite{ehnos}, and the QCD axion discussed in this paper. 

The invisible axion~\cite{Kim:1979if,Zhitnitsky:1980tq} is made invisible by presuming that the scale
associated with its decay constant is large, perhaps approaching the GUT scale. For values of $f_a \simeq 10^{11}$ GeV,
the axion could account for the cold dark matter of the Universe.
For $f_a \gg 10^{11}$ GeV, some fine-tuning of the misalignment angle is necessary. 
In~\cite{parres2,parres3,chh}, it was argued that if the radial component of the PQ symmetry breaking field takes a large initial value in the early universe, the oscillation and rotation of the PQ symmetry breaking field efficiently produce axions through parametric resonance and kinetic misalignment, respectively. As a result, smaller values of $f_a$ are allowed, thus expanding the viable parameter range for $f_a$.

It is likely that the evolution of the radial component
would be affected by inflation.  During inflation light scalar fields may randomly walk to large field values determined by the Hubble parameter during inflation and the scalar mass, 
$\langle S^2 \rangle \simeq H_I^4/m_S^2$. This allows us to set the initial condition for the PQ field, with subsequent evolution once inflation ends. Furthermore, depending on the size of the saxion quartic coupling (see Eq.~(\ref{boundary})), the evolution of the saxion/axion system may begin before reheating during an inflaton matter-dominated period, or for smaller quartic couplings, after reheating during a radiation-dominated period. In the former case, the evolution depends on the reheat temperature $T_R$
and in both cases it depends on the Hubble parameter during inflation $H_I$. 

\begin{figure}
\centering
\includegraphics[width=0.65\linewidth]{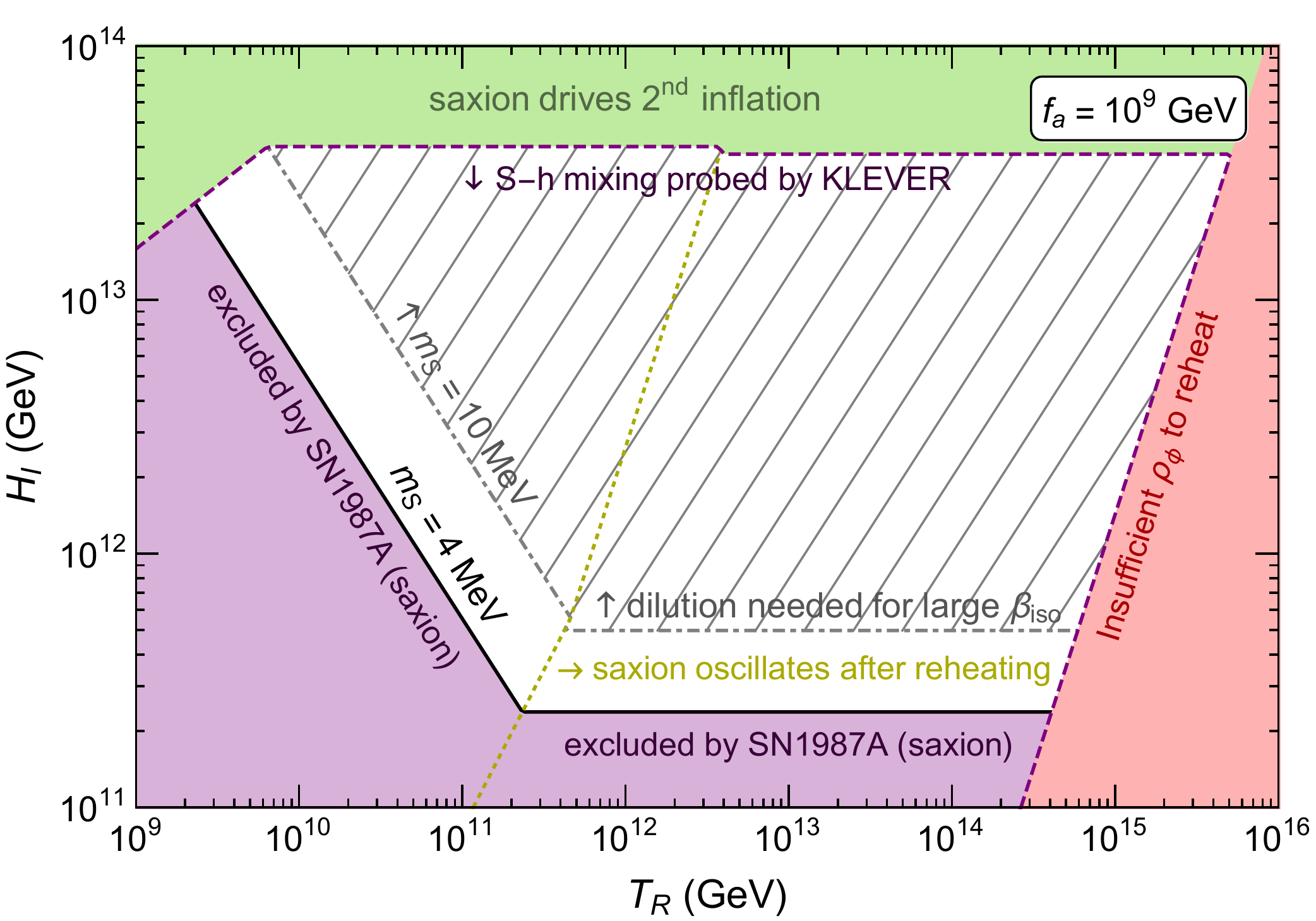}
\includegraphics[width=0.65\linewidth]{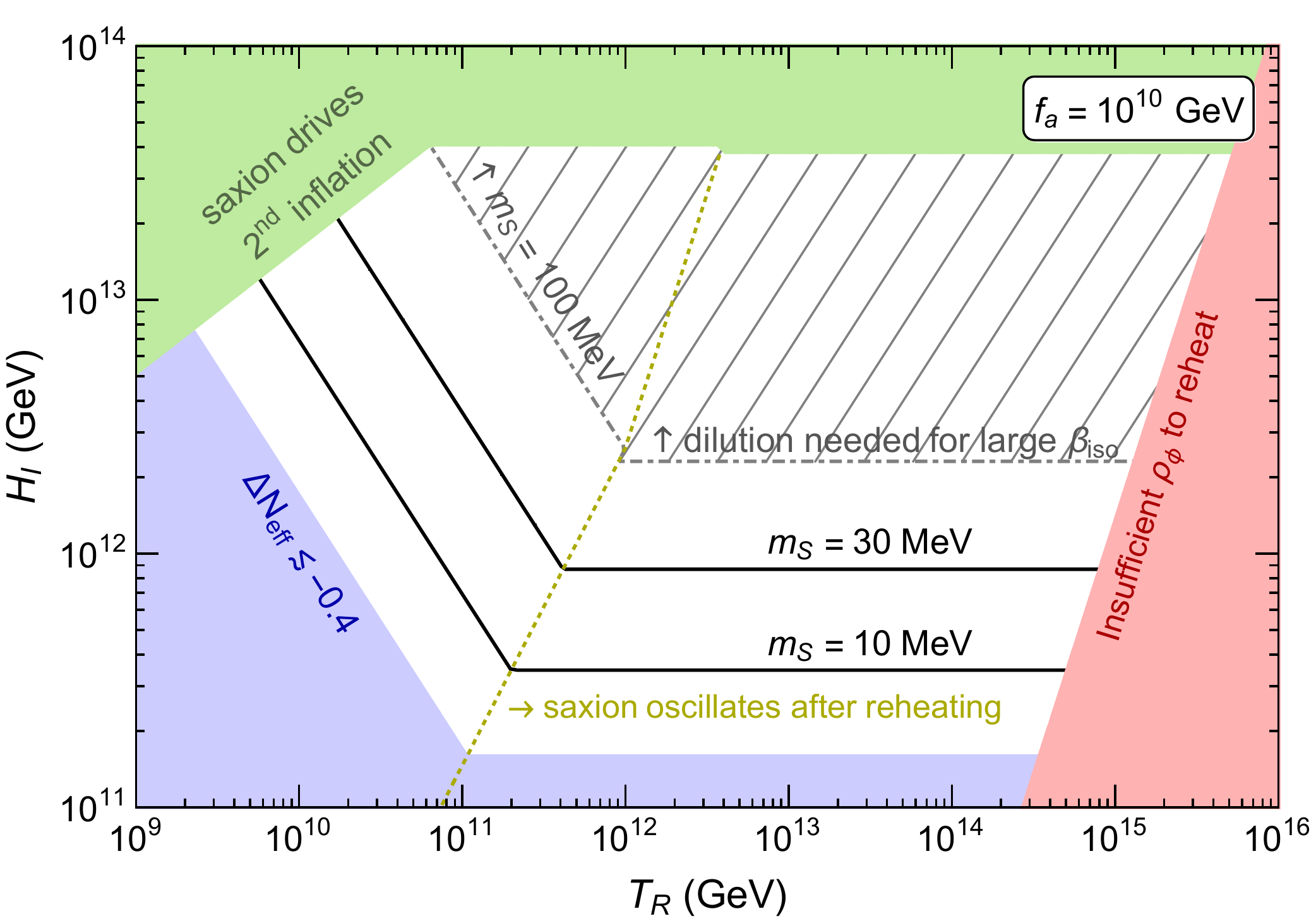}
\caption{KMM axion dark matter in the $T_R - H_I$ plane, with $\epsilon = 0.5$ and  $f_a = 10^9$ GeV (upper panel) and $f_a = 10^{10}$ GeV (lower panel). Shaded regions show
constraints due to the supernova constraint (purple), isocurvature fluctuations (gray hatched), avoidance of a second period of inflation (green), and the upper limit on $T_R$ (red).
The saxion is too light and/or its late decays conflict with BBN in the blue shaded regions. 
The gray hatched region is excluded by large isocurvature perturbations, which can be evaded by late-time dilution.
Below the purple dashed curve, the saxion-Higgs mixing nesessary to evade the supernova constraint can be probed by KLEVER.
Lines of constant
$m_S$ are shown and assume a value of $\lambda$ by  fixing the relic density of axions with $\Omega_a h^2 \simeq 0.12$. The steep  dotted line separates regions where radial oscillations begin before or after reheating.}
\label{fig:HI_TR}
\end{figure}

It is instructive therefore, to fix the axion decay
constant and consider the inflationary $T_R - H_I$ plane, as seen in Fig.~\ref{fig:HI_TR} for the case that KMM axions provide all of dark matter, with $f_a = 10^9 \, (10^{10})$ GeV in the upper (lower) panel. Shading of excluded regions is the same as in Fig.~\ref{fig:fa_TR}. For $f_a = 10^9$ GeV, none of the parameter space is allowed in the absense of a saxion-Higgs coupling.
This may be due to the lack of thermalization, saxion production in supernovae, or excessive isocurvature fluctuations.
However, if a saxion-Higgs coupling is present, the supernova constraint is evaded and the $T_R - H_I$ parameter space begins to open when $f_a \gtrsim 4 \times 10^8$ GeV.
For $f_a = 10^9$ GeV,  the range in $H_I$ is restricted between $3 \times 10^{11}$ GeV and $4 \times 10^{13}$ GeV. 

For $H_I$ near the lower end of the range (determined by the late decay of the thermalized saxion excluding $m_S < 4$~MeV), $T_R > 10^{11}$~GeV is required. The saxion mass can reach up to about 10 MeV near/inside the isocurvature limit. 
High values of $H_I$ are constrained by either avoiding a period of saxion driven inflation (at low $T_R$) or excessive isocurvature fluctuations (at high $T_R$). The isocurvature constraint can be evaded by late-time dilution; otherwise, the reheat temperature is restricted to a small range around $10^{10}$ GeV at high $H_I$.
For higher $f_a = 10^{10}$ GeV, as seen in the lower panel of Fig.~\ref{fig:HI_TR}, the constraints from late saxion decays and isocurvature fluctuations are both relaxed, though the constraint from avoiding a second period of inflation is stronger. More parameter space is allowed with a broader range in saxion masses which reach 100 MeV near the isocurvature limit. For even higher $f_a = 10^{11}$ GeV, the saxion mass can be as high as 1~GeV.

\begin{figure}
\centering
\includegraphics[width=0.65\linewidth]{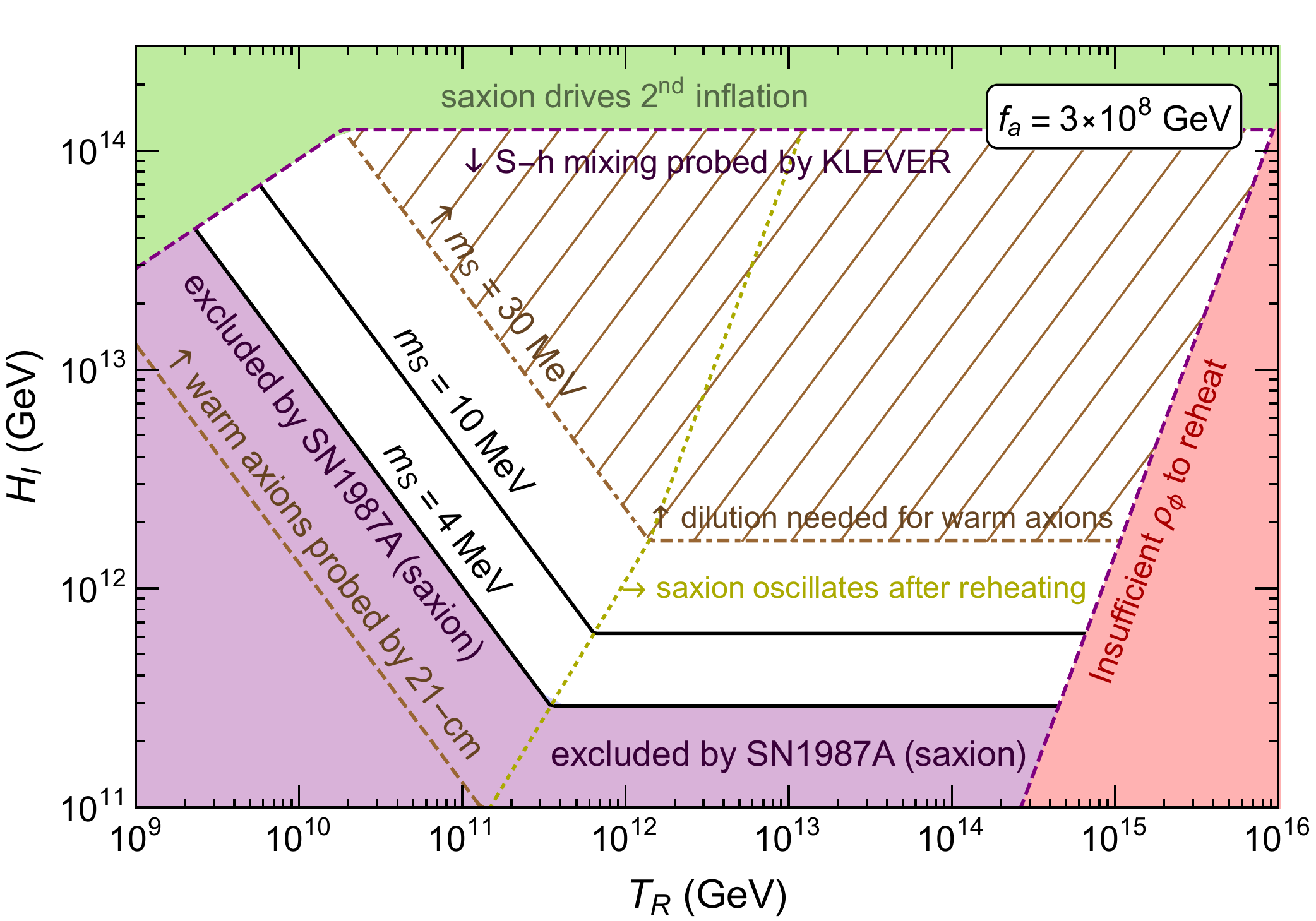}
\includegraphics[width=0.65\linewidth]{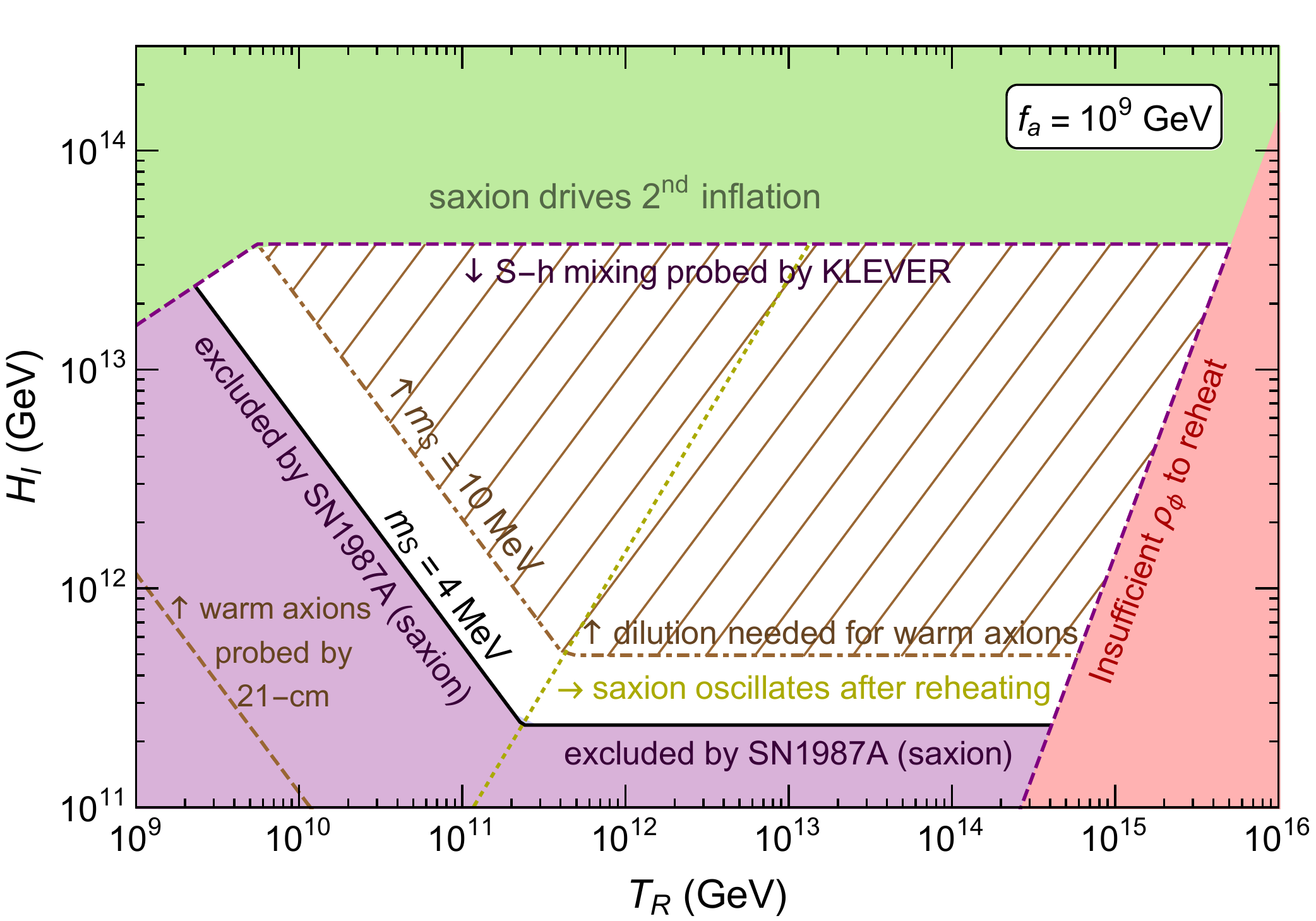}
\caption{Parametric resonance axion dark matter in the $T_R - H_I$ plane  ($\epsilon \ll 1$), with $f_a = 3 \times 10^8$ GeV (upper panel) and $f_a = 10^9$ GeV (lower panel). Green, purple and red shadings are as in Fig.~\ref{fig:HI_TR}. The brown hatched region is excluded by warm dark matter but can be opened up by late-time dilution.  The entire allowed parameter space can be probed by KLEVER (below purple dashed line), due to saxion-Higgs mixing, and by 21-cm cosmology (above brown dashed line) from warm dark matter.}
\label{fig:HI_TR_PR}
\end{figure}

In Fig. \ref{fig:HI_TR_PR}, we show the $T_R - H_I$
plane with $\epsilon \ll 1$, when the axion abundance is primarily determined by parametric resonance, for fixed values of $f_a = 3 \times 10^8$ GeV (upper panel) and $10^9$ GeV (lower panel). These results for PR axions are complementary to the KMM case shown in Fig.~\ref{fig:HI_TR}; for PR, smaller values of $f_a$ allow more parameter space. In fact, axion dark matter from PR requires a range 
$f_a = 10^8 - 2 \times 10^9$ GeV. 
Light saxions ($m_S < 4$ MeV) are excluded by the SN 1987A constraint on saxion emission. 
The warmness constraint
excludes much of the parameter space with large values of $m_S$. For  $f_a = 3 \times 10^8$ GeV and $10^9$ GeV, $m_S < 30$ MeV and $10$ MeV are required, respectively. The entire allowed parameter space for PR axion dark matter can be probed by both KLEVER and 21cm observations. 

Our scenario involves the production of topological defects, which constrains possible cosmological evolution. 1) When the kick to the angular direction is sufficiently small, the PQ field oscillations pass near the origin producing fluctuations in the field via parametric resonance~\cite{parres1}. The resulting non-thermal restoration of PQ symmetry after inflation leads to the formation of cosmic strings and domain walls~\cite{PR_DW}. For the quartic potential, this occurs for $\epsilon < 0.8$. This case is viable only if the domain wall number is unity. If $\epsilon$ is $\mathcal{O}(1)$ or the axions produced by parametric resonance are thermalized, the KMM production of axions dominates. When $\epsilon \ll 1$ and the axions are not thermalized, axion production by parametric resonance dominates. 2) A sufficiently large kick to the angular direction prevents parametric resonance and the subsequent symmetry restoration. For the quartic potential, this requires $\epsilon > 0.8$. However, in this case, fluctuations of the angular velocity lead to the production of domain walls without strings, which overclose the universe even if the domain wall number is unity. We find a strong upper bound on the Hubble scale during inflation, which excludes $\epsilon > 0.8$ with a quartic potential.

In a scenario where the radial component is fixed at the present value $f_a$ during inflation, isocurvature fluctuations give a strong upper bound on $H_I$~\cite{isoc}. The bound may be relaxed if the radial component takes a large value during inflation~\cite{Linde:1991km}. Excessively large field values however lead to parametric resonance and the formation of topological defects, giving an upper bound $H_I \lesssim 10^{12}$ GeV for simple potentials of the radial direction~\cite{HI_iso}. (See however~\cite{Harigaya:2015hha}.) The kick from the explicit PQ symmetry breaking may prevent parametric resonance and relax the upper bound on $H_I$. As stated above, this scenario does not work for the quartic potential with the initial condition determined by quantum fluctuations. Other types of potentials and/or initial conditions may render this scenario viable. 

In models for kinetic misalignment and parametric resonance, the potential of the PQ symmetry breaking field must be very flat so that the oscillating condensate has sufficient PQ charge or saxion number density. For the quartic potential, this results in extremely small coupling $\lambda^2$, so one might question whether progress has been made in solving the strong CP problem.  We view the quartic potential as a toy model; in a supersymmetric theory, the potential may be set by a supersymmetry breaking soft mass, so that the flatness is natural. Furthermore, it may be that the potential is sufficiently flat that during inflation the PQ field fluctuates to large enough values that the initial impulses for $\dot{S}$ and $\dot{\theta}$ are both dominated by an explicit PQ symmetry breaking term, from which $\epsilon \sim 1$ follows, leading to effective kinetic misalignment.

In summary, we have explored parametric resonance and kinetic misalignment in a quartic potential in the framework of general inflation models. The parameter space of $f_a$ and $m_S$ is significantly restricted compared to an unconstrained initial condition for $S_i$~\cite{parres2,chh}; in particular $f_a \sim (10^9 - 10^{11})$ GeV and $m_S = (4 - 1500)$ MeV.
For $\epsilon \ll 1$, values of $f_a$ range from $(10^8 - 2\times 10^9)$ GeV and $m_S= (4 - 100)$ MeV, where the upper end of the range depends on the value of $H_I$.
Furthermore, the range for inflationary parameters is also restricted, with $H_I \gtrsim 10^{11}$ GeV and  $ T_R> 10^9$ GeV.
For specific models of inflation the parameter space is reduced further, with $T_R = (10^9 - 10^{12})$ GeV in the Starobinsky model. This narrowing of parameters enhances the observational tests of the theory, including possible signals in BBN, supernova cooling, isocurvature density perturbations and, most importantly, a search for axions with $f_a \sim (10^9 - 10^{11})$ GeV. 

\section*{Acknowledgements}
We would like to thank G. Raffelt for helpful correspondence. The work was supported in part by the DoE Early Career Grant DE-SC0019225 (R.C.), the DoE grants DE-AC02-05CH11231 (L.H.) and DE-SC0009988 (K.H.), the NSF grant NSF-1638509 (L.H.), as well as the Raymond and Beverly Sackler Foundation Fund (K.H.). The work of K.A.O. was supported in part by DOE grant DE-SC0011842 
at the University of Minnesota. 

\appendix

\section{Critical PQ charge asymmetry}
\label{app:crit}
In the standard misalignment mechanism the axion field starts oscillating when $3H(T_{*}) \sim m_a(T_{*})$. This occurs when the Universe is radiation-dominated, where
\begin{equation}
\label{der9}
 H^2(T_{*}) = \frac{\rho_R(T_{*})}{3M_{\rm Pl}^2} = \frac{\pi^2}{90} g_* \frac{T_{*}^4}{M_{\rm Pl}^2}.
\end{equation}
Here $T_{*}$ is the temperature when the axion oscillations begin and $g_*$ is the effective degrees of freedom.
From Eq.~(\ref{der9}), we find the following axion mass when the oscillations begin
\begin{equation}
\label{der11}
m_a(T_{*}) \simeq  \frac{\sqrt{g_*} \pi  T_{*}^2}{\sqrt{10} M_{\rm Pl}}.
\end{equation}
Next, the critical PQ charge density, determined when the kinetic energy, $f_a^2 {\dot \theta}^2/2 = n_\theta^2/2f_a^2$ is equal to the barrier height, $2 m_a^2 f_a^2$ and   is given by
\begin{equation}
\label{der12}
n_{\theta,{\rm crit}} \simeq 2 m_a(T_{*}) f_a^2.
\end{equation}
By normalizing this with the entropy density
\begin{equation}
\label{der13}
s(T_{*}) \simeq \frac{2 \pi^2}{45}g_* T_{*}^3,
\end{equation}
we obtain the critical yield of the PQ charge
\begin{equation}
\label{der14}
Y_{\rm crit} \equiv \frac{n_{\theta,{\rm crit}}(T_{*})}{s(T_{*})} = \frac{9 \sqrt{5} f_a^2}{\sqrt{2g_*}\pi M_{\rm Pl} T_{*}}.
\end{equation}
When the temperature is above the QCD scale $\Lambda_{\rm QCD}$, the axion mass is given by Eq.~(\ref{eq:maT}). From that axion mass and Eq.~(\ref{der11}), we obtain
\begin{equation}
\label{der17}
T_{*} \simeq \left(\frac{m_a M_{\rm Pl}\Lambda_{\rm QCD}^4}{\pi g_*^{1/2}} \right)^{1/6}.
\end{equation}
By combining this with Eq.~(\ref{der14}), we find
\begin{equation}
\label{der19}
Y_{\rm crit} \simeq 0.11 \left( \frac{f_a}{10^{9} \, \text{GeV}}\right)^{13/6} \left(\frac{150 \, \text{MeV}}{\Lambda_{\rm QCD}} \right)^{2/3} \left( \frac{26}{g_*} \right)^{5/12} .
\end{equation}	

\section{Rotation by explicit PQ symmetry breaking}
\label{app:asym}
We consider the PQ-preserving potential $V_0$ in Eq.~(\ref{pot}) and the PQ-breaking higher dimensional potential in Eq.~(\ref{potA}). The equation of motion of $P$ is
\begin{align}
\ddot{P} + 3H \dot{P} +\frac{\partial V_{0}}{\partial P^*} + \frac{\partial V_{A}}{\partial P^*} = 0.
\end{align}
By multiplying this by $P^*$ and subtracting the complex conjugation, we obtain
\begin{align}
\ddot{\theta} + 2 \frac{\dot{S}}{S} \dot{\theta} + 3 H \dot{\theta} = \frac{i}{S^2} \left(P^* \frac{\partial V_{A}}{\partial P^*} - P \frac{\partial V_{A}}{\partial P} \right).
\end{align}
Since $S$ decreases by cosmic expansion, the right hand side of the equation is effective only for a time period $\sim m_{S_i}^{-1}$ after the beginning of oscillations. The PQ charge density right after the beginning of oscillations is
\begin{align}
n_{\theta_i} = S^2 \dot{\theta} \simeq  \frac{i}{m_{S_i}}\left(P^* \frac{\partial V_{A}}{\partial P^*} - P \frac{\partial V_{A}}{\partial P} \right)
=   \frac{1}{m_{S_i}} \left(\frac{i A}{M_{\rm Pl}^{n-3}}\right) (P^{n*} -  P^{n}).
\end{align} 
Using the decomposition $P = |P| e^{i \theta}$, we can express $n_{\theta_i}$ as
\begin{equation}
\label{der6}
n_{\theta_i} = \frac{2A |P_i|^{n} \sin{n \theta_i}}{m_{S_i} M_{\rm Pl}^{n-3}},
\end{equation}
where $\theta_i$ is the initial angle at the beginning of angular rotation/radial oscillation.
By comparing Eq.~(\ref{numden2}) with Eq.~(\ref{der6}), we obtain
\begin{equation}
\label{der8}
\epsilon = \frac{2A|P_i|^{n} \sin{n \theta_i}}{V_0(P_i) M_{\rm Pl}^{n-3}}.
\end{equation}

\section{Power-law growth of the axion fluctuation}
\label{app:dtheta}

In KMM, the PQ symmetry breaking field $P$ rotates. Let us take the time slice where the temperature of the universe is homogeneous, so that the axion potential from the QCD dynamics evolves homogeneously in the time slice. Because of the isocurvarure perturbation, different points in the universe have slightly different initial angular velocities, and the phase of $P$ evolves differently. If the difference of the phase becomes larger than $\mathcal{O}(1)$, domain walls without strings are formed after the QCD phase transition and eventually dominate the energy density of the universe. In this appendix we derive the upper bound on $H_I$.

Since we are interested in the case where the rotation remains coherent, we assume that the rotation is nearly circular.
The PQ charge conservation requires that $\dot{\theta} S^2 \propto R^{-3}$. For $S \gg f_a$, since $S\propto R^{-1}$, the angular velocity $\dot{\theta} \propto R^{-1}$.

Let us first consider the case where $P$ begins rotation after the completion of reheating. The scaling of the angular velocity is
\begin{align}
\dot{\theta} = \dot{\theta}_i \times \left(\frac{t_i}{t}\right)^{1/2},
\end{align}
where $\dot{\theta}_i$ is the angular velocity when the rotation begins at the time $t_i$.  For nearly circular motion, $\epsilon\sim 1$ and $\dot{\theta}_i t_i \sim \epsilon m_{S_i} /{H_i} \sim 1$.
The phase of $P$ is
\begin{align}
\theta = \int\dot{\theta}{\rm d}t \simeq \dot{\theta}_i t_i \left( \frac{t}{t_i} \right)^{1/2} = \dot{\theta}_i t_i \frac{S_i}{S}.
\end{align}

The fluctuation of the phase is then given by
\begin{align}
\Delta \theta \simeq \frac{\Delta \dot{\theta}_i}{\dot{\theta}_i} \dot{\theta}_i t_i \frac{S_i}{S} \simeq \frac{\Delta Y_\theta}{Y_\theta} \frac{S_i}{S} .
\end{align}
This scaling is valid until $S$ reaches $f_a$. If $\Delta \theta < \mathcal{O}(1)$ at $S = f_a$, $\Delta \theta$ does not exceed $\mathcal{O}(1)$ afterward, as can be confirmed by the scaling $\dot{\theta} \propto R^{-3}$ for $S = f_a$. Thus, the condition under which domain walls are not formed is
\begin{align}
\frac{\Delta Y_\theta}{Y_\theta} \frac{S_i}{f_a} < 1.
\end{align}
The fluctuation of $Y_\theta$ is dominantly given by that of $S_i$, ${\Delta Y_\theta}/{Y_\theta}\simeq n H_I / (2\pi S_i)$. The bound on $H_I$ is given by
\begin{align}
n \frac{H_I}{2\pi} < f_a. 
\end{align}

When $P$ begins rotation before the completion of reheating, for $S>f_a$,
\begin{align}
\theta \simeq \dot{\theta}_R t_R \frac{S_R}{S} = \dot{\theta}_i t_i \frac{S_i}{S}  \frac{\dot{\theta}_R t_R S_R}{\dot{\theta}_i t_i S_i} = \dot{\theta}_i t_i \frac{S_i}{S}  \left( \frac{H_R}{H_i}\right)^{1/3} = \dot{\theta}_i t_i \frac{S_i}{S} \left( \frac{\pi^2}{30}g_* \right)^{1/6}\left( \frac{T_R^2}{\lambda S_i M_{\rm Pl}} \right)^{1/3}.
\end{align}
We then find the condition
\begin{align}
n \frac{H_I}{2\pi} <
f_a \times \left( \frac{\pi^2}{30}g_* \right)^{-1/6}\left( \frac{\lambda S_i M_{\rm pl}}{T_R^2} \right)^{1/3} \simeq
f_a \times \left( \frac{\sqrt{2}\pi^3}{15}g_* \right)^{-1/6} \left( \frac{\lambda^{1/2} H_I M_{\rm Pl}}{T_R^2} \right)^{1/3},
\end{align}
where in the last equality we use Eq.~(\ref{si}) to fix $S_i$.

\end{document}